\documentclass[sigconf]{acmart}
\usepackage[english]{babel}
\usepackage{blindtext}
\usepackage{balance}

\usepackage{url}

\usepackage[T1]{fontenc}
\usepackage[utf8]{inputenc}
\usepackage{graphicx}
\usepackage{cleveref}
\usepackage{adjustbox}
\usepackage{caption,subcaption}
\usepackage{xspace}
\usepackage{xcolor}
\usepackage{fdsymbol}
\usepackage{wrapfig}
\usepackage{afterpage}
\usepackage{soul}
\usepackage[absolute,showboxes]{textpos}

\usepackage{verbatim} 
\usepackage{multirow} 
\usepackage{pifont}
\newcommand{\cmark}{\ding{51}}%
\newcommand{\xmark}{\ding{55}}%

\newcommand{\dg}[1]{\textcolor{black}{#1}}

\newcommand*{\eg}{\emph{e.g.,}\@\xspace}
\newcommand*{\etal}{\emph{et al.}\@\xspace}
\newcommand*{\ie}{\emph{i.e.,}\@\xspace}
\newcommand*{\viz}{\emph{viz.}\@\xspace}
\newcommand*{\etc}{\emph{etc.}\@\xspace}



\copyrightyear{2023}
\acmYear{2023}
\setcopyright{acmlicensed}
\acmConference[IMC '23]{Proceedings of the 2023 ACM Internet Measurement Conference}{October 24--26, 2023}{Montreal, QC, Canada}
\acmBooktitle{Proceedings of the 2023 ACM Internet Measurement Conference (IMC '23), October 24--26, 2023, Montreal, QC, Canada}
\acmPrice{15.00}
\acmDOI{10.1145/3618257.3624817}
\acmISBN{979-8-4007-0382-9/23/10}

\begin{document}

\setlength{\TPHorizModule}{\paperwidth}
\setlength{\TPVertModule}{\paperheight}
\TPMargin{5pt}
\begin{textblock}{0.8}(0.1,0.02)
     \noindent
     \footnotesize
     If you cite this paper, please use the IMC reference:
     Zeya Umayya, Dhruv Malik, Devashish Gosain, and Piyush Kumar Sharma. PTPerf: On the performance evaluation of Tor Pluggable Transports. In Proceedings of the ACM Internet Measurement Conference (IMC ’23), October 24–26, 2023, Montreal, QC, Canada.
\end{textblock} 

\title{PTPerf: On the performance evaluation of Tor Pluggable Transports}

\author{Zeya Umayya}
    \affiliation{%
        \institution{IIIT Delhi} 
        \country{}
}
\email{zeyau@iiitd.ac.in}
\author{Dhruv Malik}
    \affiliation{%
        \institution{IIIT Delhi}
        \country{}
}
\email{dhruv20373@iiitd.ac.in }
\author{Devashish Gosain}
    \affiliation{%
        \institution{imec-COSIC, KU Leuven}
        \institution{BITS Pilani Goa}
        \country{}
}
\email{devashishg@goa.bits-pilani.ac.in}
\author{Piyush Kumar Sharma}
    \affiliation{%
         \institution{imec-COSIC, KU Leuven}
         \country{}
}
\email{pkumar@esat.kuleuven.be}

\begin{abstract}

Tor, one of the most popular censorship circumvention systems, faces regular blocking attempts by censors. 
Thus, to facilitate access, it relies on ``pluggable transports'' (PTs) that disguise Tor's traffic and make it hard for the adversary to block Tor. 
However, these are not yet well studied and compared for the performance they provide to the users. 
Thus, we conduct a first comparative performance evaluation of a total of 12 PTs---the ones currently supported by the Tor project and those that can be integrated in the future. 

Our results reveal multiple facets of the PT ecosystem. 
\textbf{(1)} PTs' \textbf{download time} significantly varies even under similar network conditions. 
\textbf{(2)} All PTs are not equally \textbf{reliable}.
Thus, clients who regularly suffer censorship may falsely believe that such PTs are blocked.
\textbf{(3)} PT performance depends on the underlying \textbf{communication primitive}.
\textbf{(4)} PTs performance significantly depends on the website \textbf{access method} (browser or command-line). Surprisingly, for some PTs, website access time was even less than vanilla Tor.

Based on our findings from more than $1.25$M measurements, we provide recommendations about selecting PTs
and believe that our study can facilitate access for users who face censorship.

\end{abstract}

\maketitle

\section{Introduction}

Internet censorship has become pervasive over the years, and hence, there exists a plethora of research, both in identifying censorship mechanisms~\cite{ensafi2015examining,wang2017your, sundara2020censored, yadav2018light, niaki2020iclab, ramesh2020decentralized} as well as bypassing them~\cite{dingledine2004tor, nasr2020massbrowser, vandersloot2020running}. Among various circumvention solutions, Tor is one of the oldest and most popular systems~\cite{dingledine2004tor} with more than 2.5M average concurrent users~\cite{david_tor_users}. Thus, Tor has witnessed several attempts by adversaries to prevent users from accessing it \cite{winter2012great, dunna2018analyzing, ensafi2015examining}. But, such attempts have been dealt with a suitable workaround, leading to a cat-and-mouse game between the censors and anti-censorship researchers. 

Thus, it is becoming increasingly important to enhance the circumvention ecosystem. This is also highlighted by Russia's excessive Internet restrictions during the Ukraine war~\cite{Russia-ban-net, Russia-ban-net2}, which prevented Russian citizens from accessing information from outside the country. Similarly, in September 2022, civil unrest and massive protests erupted in Iran \cite{IR-protest-UN}. To curb the unrest, the Iranian government blocked access to numerous Internet services \cite{IR-blocks-Internet-Forbes, IR-blocks-Internet-Guardian}, along with regular Internet shutdowns \cite{IR-blocks-net-shut-1, IR-blocks-net-shut-2}. 
The regime also censored the Tor network \cite{IR-block-Tor}. Thus, people resorted to using ``pluggable transports'' (PTs) offered by the Tor project \cite{IR-snow-users} to access the blocked content.

PTs essentially work as a gateway to the Tor network and masquerade Tor traffic as some random-looking or benign Internet application traffic (\eg web streaming). 
This makes it hard for the adversary to fingerprint and filter Tor traffic. 
Thus, almost all research on PTs focuses on making them unobservable to the censor to resist blocking.
However, another crucial aspect that warrants attention is the \textit{performance} provided by these PTs when used in conjunction with Tor. In scenarios where users are stressed and under fear of using circumvention solutions, poorly performing PTs might add to their woes. There exist many PTs, but there is no comprehensive analysis about how they perform, how reliable they are, whether they can facilitate access only for websites or for other applications like bulk downloads \etc

To this end, we conduct the first comparative performance evaluation of PTs. We evaluated a total of 12 PTs---meek \cite{meek}, conjure \cite{torConjure}, snowflake \cite{snowflake}, dnstt \cite{dnstt}, webtunnel \cite{webTunnel}, marionette \cite{dyer2015marionette}, psiphon \cite{psiphon}, stegotorus \cite{weinberg2012stegotorus}, camoufler \cite{sharma2021camoufler}, cloak \cite{cloak}, shadowsocks \cite{shadowsocks}, and obfs4 \cite{obfs4}. We assess their performance for the most common use cases of accessing websites and bulk downloads. Apart from website access (or file download), we also consider additional parameters such as the time to first byte (TTFB), speed index \cite{speedIndex}, the impact of location, the impact of the transmission medium, and how reliably the PTs can download the content.
Through 1.25M measurements spanning more than a year, we made multiple observations and deduced statistically significant inferences.
\vspace{1mm}

\noindent \textbf{Performance trend for website accesses:}
For website access performance, we visited the Tranco top 1k and a curated list of blocked 1k websites. Our evaluation reveals that obfs4, webtunnel, cloak, and conjure were the best-performing PTs with an average access time of $2.4 s$, $3.2 s$, $2.8 s$, and $2.5 s$, respectively. But, camoufler, meek, and dnstt took significantly more time for the same with $12.8s$, $5.8s$, and $4.4s$. The least performing PT with respect to website access was marionette ($20.8s$), with $8\times$ more time in comparison to vanilla Tor ($2.3s$).
\vspace{1mm}

\noindent \textbf{Performance trend for bulk downloads and reliability implications:} For bulk downloads, we recorded the time for downloading files of varying sizes (5, 10, 20, 50, and 100 MB). Here again, we observed that some PTs, \eg obfs4 and cloak, performed better than all other PTs. For instance, they took 64s and 53s to download a file of 50 MB; whereas camoufler incurred the maximum time of $173s$ to download the same file. 
Interestingly, we observed that some PTs were not able to download the files in all the download attempts. We quantified this unreliable behavior and observed that meek, dnstt, and snowflake were unable to download files successfully in more than 80\% of the attempts. Thus, these PTs may not be the best choice for downloading bulk content.
\vspace{2mm}

\noindent \textbf{Impact of change in location and transmission medium:} We performed our experiments (website access and bulk downloads) across nine different client and server locations. We did not observe any change in \textit{trends} for the performance of these PTs based on location. Similarly, our experiments of accessing these PTs via wireless medium did not have any significant impact on performance. Thus, any of these PTs can be used across locations and across mediums without any observable degradation in performance. 
\vspace{1mm}

\noindent \textbf{Inferences on overall PT ecosystem:} Using our measurement results, we found multiple insights. \textit{Firstly,} we analyzed the performance of these PTs based on their underlying technology that provides unobservability to the PTs. To this end, we categorize them as tunneling, mimicry, fully encrypted, and proxy-layer PTs (for details, see \Cref{sec:background}). We observed that the fully-encrypted and proxy-layer-based PTs performed better than the other categories. 
This is because the PTs in mimicry and tunneling categories are restricted by the underlying protocol they try to mimic or tunnel. We observed variations in performance even within the categories. For example, in tunneling-based PTs, camoufler and dnstt took significantly longer time to access resources in comparison to webtunnel. This is because camoufler relies on instant messaging apps to send content and is thus rate-limited in the number of requests a client can send using the API of these apps. Similarly, dnstt requires sending DNS requests to DoH/DoT servers for which the response size is limited, resulting in slightly reduced performance. However, there is no such limitation in webtunnel. We discuss other categories and their performance in detail in \Cref{eval:web}.

\textit{Secondly,} as a common practice for evaluating circumvention systems, we used a command-line utility \texttt{curl} to perform our experiments. However, an actual user generally accesses websites using the browser; thus, we did additional experiments to access websites via browser automation using the selenium framework \cite{selenium}. As expected, we observed that overall, all the PTs took more time to access the websites compared to \texttt{curl} as multiple web resources would be requested via the browser compared to the default page of the website requested by \texttt{curl}. However, with the selenium-based website access, we noticed that a few PTs \viz obfs4, webtunnel, and conjure incurred less time than vanilla Tor. We investigated this observation and found that the first hop (in the Tor circuit construction) may significantly impact the performance observed via these PTs (see details in \Cref{PTs-better-than-Tor}).

\textit{Thirdly,} some of our measurements coincided with the unrest in Iran \cite{IR-blocks-Internet-Forbes, IR-blocks-Internet-Guardian}. For this duration, there was a \dg{sudden} increase in the usage of PTs. We particularly were able to see and quantify variation in the performance of snowflake during this time as it was the most widely used PT during the unrest \dg{in Iran} \cite{snow-users-by-country}. Our results indicate that under high load, the performance of these PTs deteriorates, and further efforts are required to scale these systems during the peak loads (see \Cref{subsec:snowflake-load-iran}).

Since our measurement study involved accessing websites and downloading files via the public Tor network, we carefully planned our measurements so as not to overload the volunteer-operated Tor relays. We followed the ethical practices mentioned in the Belmont report \cite{beauchamp2008belmont} throughout the study (see \Cref{sec:ethics}).

Notably, we analyzed a total of 28 PTs for evaluation, out of which only 12 were either already integrated or had the possibility of integration with Tor. We provide a summary of all these 28 PTs along with their features and implementation challenges in Table \ref{table:allpt}.
We make our code and analysis scripts public at \cite{ptperf_zenodo, ptperf_github}.

\section{Background}
\label{sec:background}

Tor is an overlay network of proxies (or relays) that uses onion routing to facilitate \textit{anonymous} communication over the Internet \cite{dingledine2004tor}. Since Tor is a proxy-based system, it is often also used to bypass censorship in various countries. This results in censors restricting access to Tor by blocking the publicly known IP addresses of relays~\cite{lewman2012tor,dingledine2012governments}. Tor thus relies on \textit{pluggable transports} (PTs) to safeguard itself from being blocked by determined censors such as China. These PTs facilitate access to the Tor network by providing alternate means to connect. They disguise the Tor traffic, making them difficult to classify and block. Interestingly, some existing anti-censorship systems (\eg camoufler \cite{sharma2021camoufler}, psiphon \cite{psiphon}) can also be used to transport the Tor traffic outside the censor's purview. Thus, in this work, we test the feasibility of using them as PTs in addition to the standard PTs supported by Tor (\eg obfs4 \cite{obfs4}).

Generally, a PT consists of two components: the client and the server (proxy). The client connects to the server while maintaining the properties (such as traffic patterns) of the corresponding transport. Once the connection is established, it acts as a tunnel to transfer Tor traffic.

There are multiple ways in which pluggable transports can be categorized. For example, they can be classified based on--- \textit{communication primitive} they use to evade blocking (mimicry, obfuscation, adding an extra layer of the proxy, tunneling \etc)---their \textit{integration status} with Tor (officially integrated, in the process of integration \etc)---their \textit{implementation style} (PT server acting as Tor guard, PT server separate from Tor guard \etc). 
Since PTs use different communication primitives to masquerade their clients' traffic as uncensored traffic, it may impact their performance. For instance, dnstt \cite{dnstt} sends content inside the encrypted DNS request-response pairs using DNS over HTTPS servers \cite{dohr, dotr}, whereas cloak \cite{cloak} mimics users' traffic to resemble regular web browser traffic.
Thus, we categorize PTs based on the underlying technology they use to evade blocking.

\subsection{Proxy-layer based pluggable transports} 

This class of PTs adds an additional layer of proxy before connecting with the Tor network. 

\textbf{\textit{Meek}} \cite{meek}, uses domain fronting \cite{fifield2015blocking} as the underlying technology to evade censors. Domain fronting enables a user to deploy a service (\eg \texttt{appspot.com}) with a service-specific domain (\eg \texttt{forbidden.appspot.com}). However, to access the deployed service, the user can use the domain of the fronting service (\eg \texttt{appspot.com}) in the plain text headers and specify the service-specific domain in the encrypted payload. This keeps a pretense to the adversary that the client is accessing some benign domain of the fronting service (visible in all plain-text fields \viz DNS and TLS ClientHello SNI field). However, in practice, the deployed service facilitates the client to access the censored content, all hidden inside the encrypted HTTPS communication.\footnote{Note that assigning PTs to different categories is not mutually exclusive. For instance, meek can also be considered as tunneling as it tunnels the censored traffic inside the innocuous-looking web traffic. As a best effort, we assign a PT to a category where the underlying technology could heavily impact its performance.}

\textbf{\textit{Psiphon}}~\cite{psiphon} provides a network of proxy servers that the clients can connect to circumvent censorship. The core mechanism for connecting to psiphon servers is establishing an SSH tunnel~\cite{psiphon-github}. The SSH public key for authenticating to the server is pre-shared with the client. Psiphon can also be manually configured to add different tunneling or obfuscation protocols. For the purpose of evaluation, we use the default psiphon configuration that uses SSH tunneling.

\textbf{\textit{Conjure}} \cite{frolov2019conjure} is a
refraction networking (formerly called decoy routing) system \cite{vandersloot2020running} that relies on support from an Internet Service Provider (ISP) to deploy the circumvention infrastructure. The basic idea behind decoy routing is to make a router as a proxy (also known as a decoy router) such that when a user sends a request to an uncensored website, an on-path decoy router can proxy the request to a censored website.
To a censor, it would appear that the user is communicating with a legitimate website, while in practice, it accesses the censored website (with the help of the decoy router).
However, there are a limited number of such uncensored sites; thus, conjure leverages the unused IP address space of ISPs deploying the decoy routing infrastructure. Instead of using actual uncensored destinations for proxy connections, it connects to ISP's phantom IP addresses where no web server exists. 


\textbf{\textit{Snowflake}} \cite{snowflake} relies on WebRTC services to function. A snowflake client uses a domain fronted (or HTTPS) server (known as a broker) to connect to volunteer-run short-lived WebRTC snowflake proxies. Broker is used only once for connecting clients to the proxies. Subsequently, clients and proxies communicate directly (without involving the broker).
Snowflake relies on the availability of a large number of WebRTC proxies such that blocking all of them by the censor is difficult. Thus, these proxies are essentially browser plugins designed so non-technical users can easily run them. 
It must be noted that since the target users deploying the proxy are in a home network, they would probably be behind a NAT. Thus, it would be difficult for the censor to recklessly block requests to public IPs as they would unintentionally also block access to legitimate users behind those NATs.

\begin{figure*}[h!]
	\centering
	\includegraphics[width=0.97\textwidth]{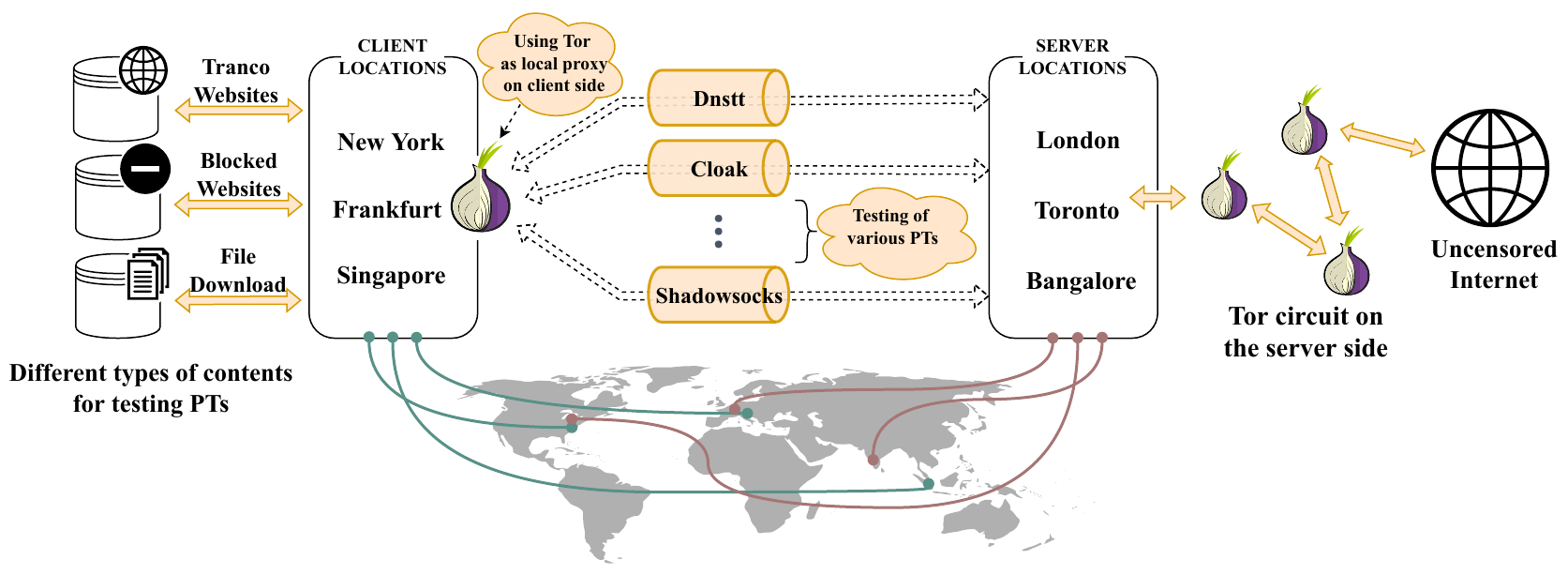}
	\caption{Overview of our approach: At multiple geographic locations, we host our pluggable transport (PT) clients and servers. We measure the performance of 12 PTs, including dnstt, cloak, and shadowsocks \etc We measure the download time for different targets \viz Tranco top-1k, 1k blocked websites, and files of different sizes.}
	\label{fig:approach}
\end{figure*}

\subsection{Tunneling-based pluggable transports}

These PTs encapsulate the web traffic in standard application protocol messages \eg video streams, IM apps \etc

\textit{\textbf{Dnstt}} \cite{dnstt} is a recent tunneling-based solution. To a censor, it would appear that the dnstt client is communicating with a DNS resolver over an encrypted channel. But in practice, the client would fetch the censored content. It takes advantage of DNS over HTTPS/TLS resolvers \cite{dohr, dotr} to hide the traffic from a censor. 
These recursive resolvers (in our case OpenDNS DoH resolvers \cite{dnstt_doh_server}) act as a proxy to forward the dnstt client's requests to dnstt servers.
Note that there is a limit on the size of response messages that a public DoH server supports, typically 512 bytes \cite{dnstt}.

\textit{\textbf{Camoufler}} \cite{sharma2021camoufler} uses instant messaging (IM) applications (such as WhatsApp, Telegram \etc) as a medium to tunnel censored content. A client is required to have an account on any IM app. It can then use this account to send messages/content to another IM account in a non-censored region. The IM app in the non-censored region deploys the proxy software in the background. Thus, the censor only observes standard IM traffic being exchanged between IM clients and IM servers. Moreover, IM apps are generally end-to-end encrypted, making it more difficult for the censor to identify \textit{camoufler}. 
Thus, to block the system, the censor requires blocking all IM apps. This may cause collateral damage as IM apps are an important part of digital space with high personal and business usage.

\textit{\textbf{WebTunnel}} \cite{webTunnel} tunnels the censored content inside the regular HTTPS traffic. At its core, it uses HTTPT \cite{frolov2020httpt} (an existing anti-censorship system) to establish the HTTP tunnel. 
The webtunnel server side has two components---a customized server program (with a valid TLS certificate) 
and the Tor bridge program. The client first establishes a TLS connection with the webtunnel server, and thus a censor can only observe a regular TLS connection with an unblocked domain. Once the connection is established, the webtunnel client sends Tor traffic (received from the Tor client utility) inside the established HTTPS tunnel. On the other side, the webtunnel server decapsulates the Tor traffic and forwards the traffic to the Tor bridge utility.

\subsection{Mimicry-based pluggable transports}

These PTs disguise and transport blocked content as other regular application protocol messages. They attempt to mimic all the features of the underlying protocol.

\textit{\textbf{Cloak}} \cite{cloak} provides a communication channel that obfuscates the user traffic to resemble regular web browser traffic. It can also be used to obfuscate the proxy's traffic. The cloak server uses a series of steganographic techniques to authenticate clients in zero RTT.  
Whenever a client wants to connect to the cloak server, it creates a TLS ClientHello packet with appropriate values, including the client random, desired proxy type (\eg Tor), and a non-blocked domain name (\eg \texttt{uncensored.com}). The plain text TLS ClientHello SNI field is set as an unblocked domain to fool the adversary. Once the cloak server receives the packet from a client, it validates the packet by inspecting the client random, and on successful validation, relays the traffic to the Tor network.

\textit{\textbf{Stegotorus}} \cite{weinberg2012stegotorus} uses steganography to encode and hide the Tor clients' data into innocuous-looking cover traffic such as HTTP. Stegotorus client uses a ``chopper'' method to convert fixed-size Tor cells to variable-sized blocks. At the client side, the chopper protocol sends these blocks unordered over multiple TCP connections to the Stegotorus server that reassembles them in Tor cells format. Finally, it relays the traffic to the Tor network.

\textit{\textbf{Marionette}} \cite{dyer2015marionette} tries to obfuscate network traffic by giving a tunable and programmable network traffic generator. Marionette gives the control of using appropriate obfuscation methods to clients since each client might be facing a different kind of adversary. Such capability is controlled by using a lightweight domain-specific language. It takes arguments such as \textit{connection type} (\ie TCP), \textit{port} where the server is listening, and the \textit{method} (\eg FTP) with which Marionette obfuscates the traffic. The flexibility to program the encrypted traffic properties using statistical methods makes it difficult for the censors to block the system using standard traffic analysis.

\subsection{Fully-encrypted pluggable transports} These PTs encrypt the traffic to make it appear as a random byte stream to a censor.

\textit{\textbf{Obfs4}} \cite{obfs4} is a proxy system. It is a successor of the the scramblesuit protocol proposed by Winter \etal \cite{winter2013scramblesuit,scrambleSuit_website}. Apart from the traditional proxy functionality, obfs4 provides additional features. These features include (1) obfuscating the application data (using scramblesuit) such that it appears completely random to a censor, (2) authenticating the legitimate clients using an out-of-band shared secret, making it difficult for the censor to probe for proxy functionality.

\textit{\textbf{Shadowsocks}} \cite{shadowsocks} is also a proxy system that obfuscates the proxied content. It comprises two components. A client that encrypts the application traffic to appear as a uniformly random byte stream. A server that decrypts and proxies the client's traffic to the blocked destinations. There are multiple configurable options available to the shadowsocks clients that can lead to different types of traffic obfuscation.

\begin{table*}[t!]
    \centering
    \begin{adjustbox}{max width=\textwidth}
        \begin{tabular}{lcl}
            \toprule
            \textbf{Measurement Type}         & \textbf{Number of Measurements}    & \textbf{Target}   \\
            \midrule
                Website Download (curl)       &      149.5 k                           &    Tranco top-1k \& CBL-1k         \\
                Website Download (selenium)   &      174 k                          &    Tranco top-1k \& CBL-1k         \\ 
                File Downloads (curl)         &      2.7 k                          &    5 MB, 10 MB, 20 MB, 50 MB, 100 MB      \\
                File Downloads (selenium)     &      2.7 k                            &    5 MB, 10 MB, 20 MB, 50 MB, 100 MB      \\
                Medium Change (wired/wireless) &      60 k                           &    Tranco top-500 \& CBL-500 \\
                Speed Index                    &      60 k                           &    Tranco top-1k \\
                Pluggable Transport Overhead   &      40 k                            &    Tranco top-1k\\
                Location Variation            &      686 k                           &    Tranco top-1k \& CBL-1k       \\ 
             \bottomrule
        \end{tabular}
    \end{adjustbox}
        \caption{
        Overview of different measurement types. CBL-1k represents 1000 blocked websites from Citizen Lab~\cite{citizenlab} and Berkman research center~\cite{berkman}.
        }
    \label{tab:measurements}
\end{table*}

\subsection{Other pluggable transports}

Apart from the PTs mentioned above, there exist other circumvention systems that cannot be evaluated as potential PTs due to various reasons. Some of them have publicly available source code but still can not be integrated with Tor due to reasons like dependency and code compilation issues \eg deltaShaper.
Then, some systems like covertCast \cite{mcpherson2016covertcast} are non-functional (\eg source code not available). 
We summarize all such PTs in Table~\ref{table:allpt}.

\section{Related Work} 
\label{sec:related-work}

The prior work on pluggable transports has mainly focused on their detectability by a censor. Khattak \etal \cite{khattak2016sok,khattak2014systemization} surveyed different censorship resistance systems, including the PTs. They proposed a framework to gauge and classify the circumvention capabilities of such systems. 
Other studies \cite{shahbar2017analysis, kwan2021exploring} focus on detecting different PTs under changing adversarial conditions. The authors show that the packet size and number of bytes sent in a flow are important features for detecting pluggable transports. 
Similar detection strategies have been developed by multiple other studies \cite{he2019detection,soleimani2018real, liang2020obfs4, xu2021obfuscated} that use machine learning methods to evaluate meek, obfs3, obfs4, scrambleSuit \cite{winter2013scramblesuit} and Format-Transforming Encryption (FTE) \cite{dyer2013protocol}. 
However, none of the above studies focus on the performance evaluation of PTs with respect to real-world deployment, varying geographical locations \emph{etc}. Thus, in this research, we present the first study to quantify the comparative performance of PTs comprehensively.

\section{Evaluation \& Analysis}

In this section, we present the evaluation of the 12 PTs described in \Cref{sec:background}.  
\Cref{fig:approach} illustrates the overview of our experimental setup and approach, and Table~\ref{tab:measurements} presents an overview of the individual measurements. 
We now describe our experimental setup, performance parameters, and the evaluation of the PTs.

\subsection{Experimental setup}
\label{subsec:exp-setup}

A PT consists of two components: a PT client and a PT server. Depending upon how PTs are implemented (\eg some PT servers can act as guard nodes and some cannot), they have different configurations.
The PTs we study can be divided into three sets based on implementation.  

The first set of PTs is where the PT server also acts as the first hop in a Tor circuit. This effectively leads to a total of three hops between the client and the website---PT server/guard, middle Tor relay, and exit Tor relay. 
PTs belonging to this category are obfs4 meek, conjure, and webtunnel. 
Note that in dnstt, the PT server acts as a guard node. But still, there are four hops between the client and the website as the client first communicates with the DoH recursive resolver before accessing the Tor network.

The second set of PTs is where the PT server is separate from the Tor circuit's first hop, effectively leading to four hops between the client and the website. In this class of PTs, at the client side, application traffic is sent to the Tor client, which in turn forwards it to the PT client utility (all running inside the client's host machine). PT client utility then transfers the traffic to the PT server (a different host). 
PT server relays the traffic to the Tor guard node specified by the client. The PTs in this category include shadowsocks, snowflake, camoufler, stegotorus, massbrowser, and psiphon.

The third set of PTs also has four hops between the client and the website. However, in this category, on the client side, application traffic is directly sent to the PT client (all inside the client's host). PT client forwards the received traffic to the PT server (a different host), where the standard Tor client runs. The Tor client then creates a three-hop circuit to the website. PTs in this category include marionette and cloak.

Wherever required, we hosted our PT client, PT server, and Tor client on cloud hosting infrastructure.
Our experimental setups include a client machine that runs the PT client along with the Tor client utility (for sets 1 and 2). The other end of the setup consisted of a server machine that runs the server PT utility (and Tor client utility for set 3).
For accessing the default webpage of websites, the client requests a web resource with the help of \texttt{curl}~\cite{curl} or selenium (for browser automation). 
We configured \texttt{curl} to send all the requests to the local SOCKS port. This SOCKS port was of the Tor client utility for sets 1 and 2, whereas it was of the PT client for set 3. Based on the underlying transport, PT obfuscates and sends the content to the PT server. The PT server then forwards the request to the Tor network, through which the request finally reaches the web server that hosts the content. 
Additionally, a significant implementation effort was required to integrate some of the PTs with Tor. We highlight some of those challenges in Table~\ref{table:allpt} and \Cref{sec:implemnt}.

Using our experimental setup, we launched a measurement campaign to measure the performance of 12 PTs with different parameters, targets, location variation \textit{etc.}\footnote{For obfs4, meek, and snowflake PTs, we use their default servers (provided by Tor). For the rest, we host our own PT servers as they are currently not supported by Tor. See \Cref{sec:implemnt} for details.}
Table~\ref{tab:measurements} presents an overview of our individual measurements.

\subsection{Website access time}
\label{eval:web}
Website access time captures a prevalent use case of these PTs.
It represents the time the PTs took to access the default web pages of different websites. A PT that yields lower access time is better suited for accessing websites.

\begin{figure*}
	\centering	
 \begin{subfigure}{0.65\textwidth}
    \centering
    \includegraphics[width=\textwidth]{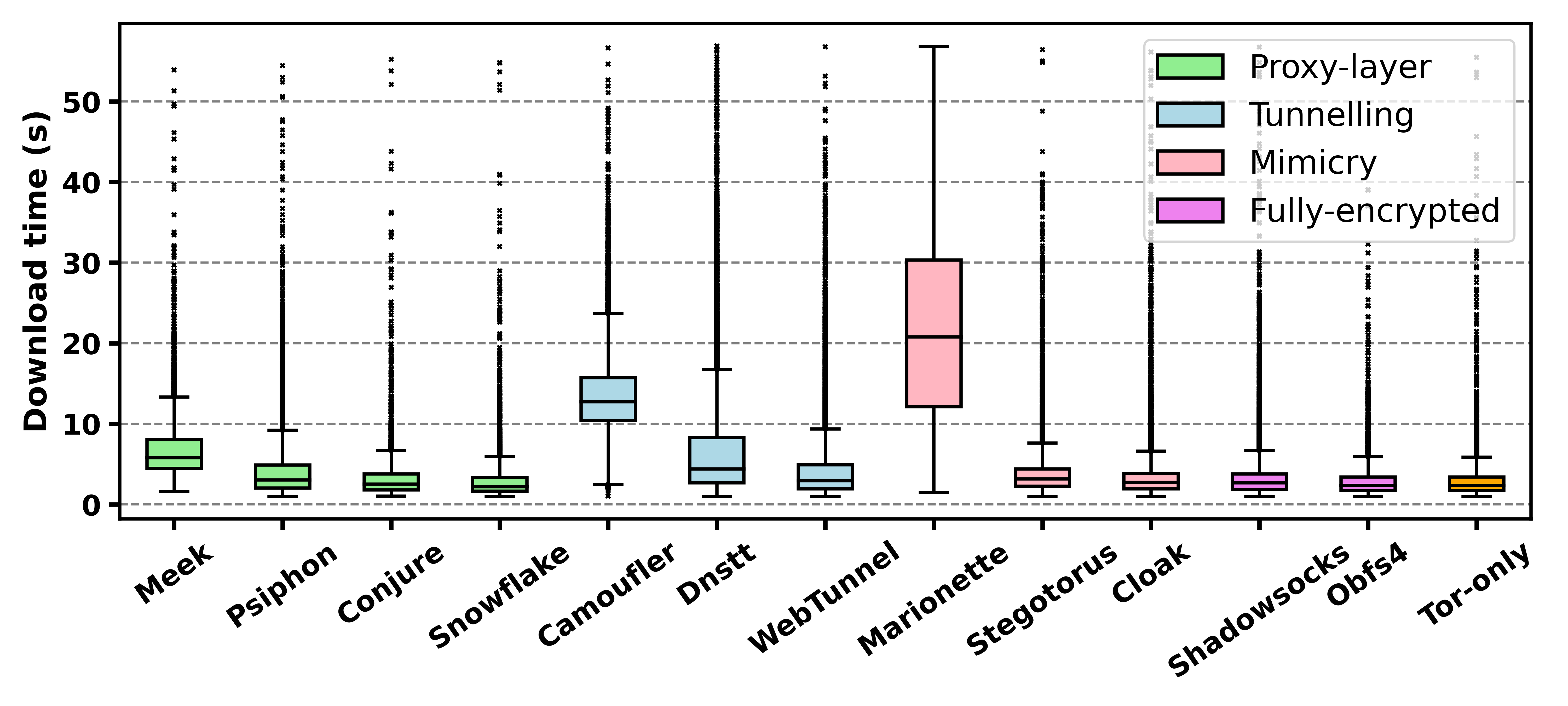}
	\caption{Curl}
	\label{fig:web_access}
\end{subfigure}

\begin{subfigure}{.65\textwidth}
	\centering	
    \includegraphics[width=\textwidth]{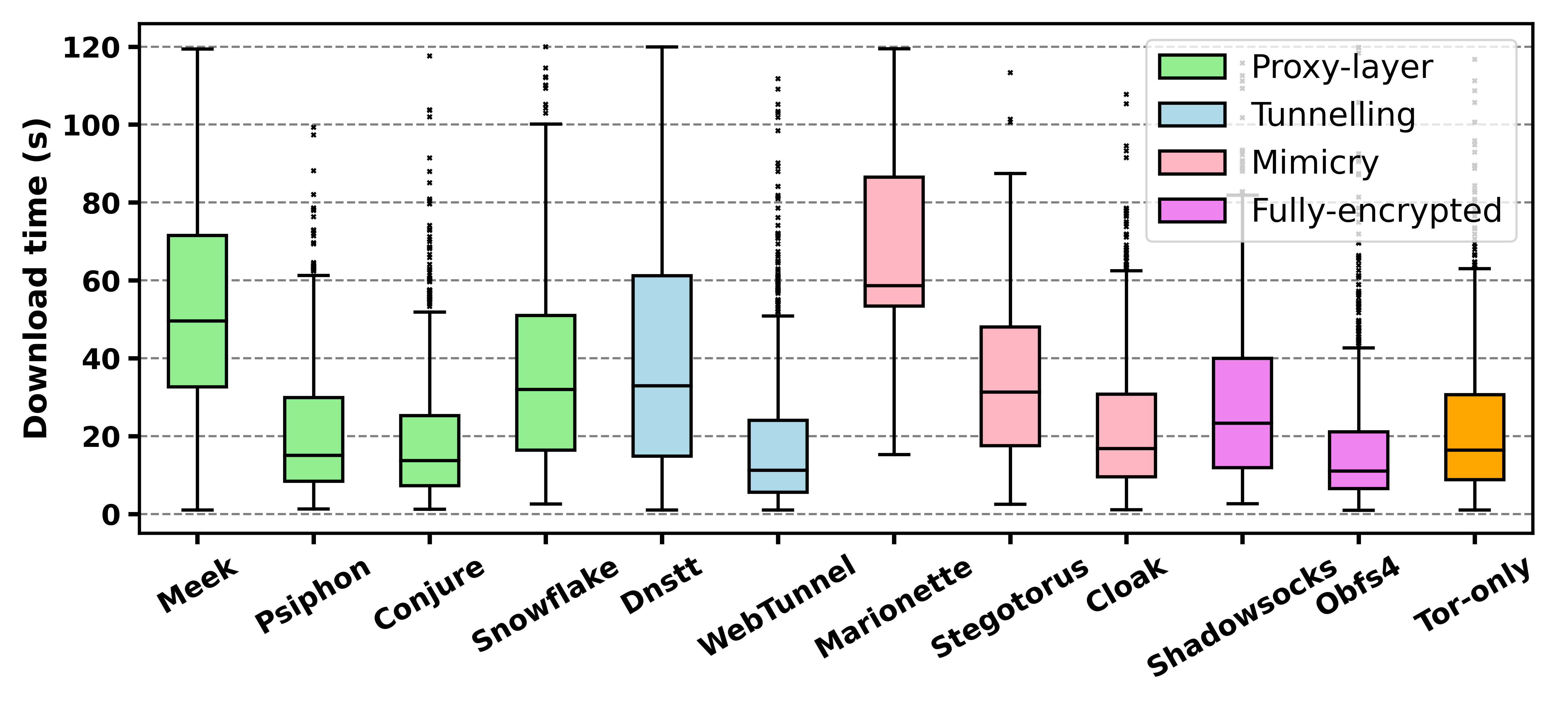}
	\caption{Selenium}
	\label{fig:web_access_selenium}
 \end{subfigure}
 \caption{Website access time for all pluggable transports.}
 \end{figure*}
 
To assess the web access performance, we consider two sets of websites: (1) \textbf{Tranco top-1k} popular websites~\cite{pochat2018tranco}, and (2) randomly selected 1000 potentially blocked websites from Citizen Lab~\cite{citizenlab} and Berkman research center~\cite{berkman} lists (abbreviated as \textbf{CBL-1k}). The CBL-1k websites are a good representative of websites that Tor clients might access.

Using \texttt{curl}, we accessed each website five times using PT's default configuration and computed the average access time per website. \Cref{fig:web_access} shows the box plot of average access time for each of Tranco top-1k and CBL-1k websites (via different PTs and vanilla Tor). The PTs are arranged based on their classification category (\eg green color representing proxy-layer PTs). Within each category, we arrange them in decreasing value of the median download time.
To measure the statistical significance of our results, we performed paired t-tests \cite{hsu2014paired} for each pair of PTs. We report the corresponding \textit{P}-value, t-value, 95\% confidence interval (CI), and the mean difference in Appendix Tables \ref{tab:Fig2a-first} and \ref{tab:Fig2a-second}.

Overall, we observe that proxy-layer and fully encrypted PTs perform better than tunneling-based and mimicry-based PTs.
The paired t-test results reveal that the mean download time of proxy-layer PTs is 2.88s less than tunneling-based PTs (t=-20.23, \textit{P}<.001) with 95\% CI [-3.16, -2.60] and 3.23s less than mimicry-based PTs (t=-24.55, \textit{P}<.001) with 95\% CI [-3.49, -2.97]. Similarly, the mean download time of fully encrypted PTs is 4.91s less than tunneling-based PTs (t=-30.26, \textit{P}<.001) with 95\% CI [-5.23, -4.59] and 5.21s less than mimicry-based PTs (t=-38.25, \textit{P}<.001) with 95\% CI [-5.48, -4.94]. We tabulate the test results for different PT category pairs in Appendix \Cref{tab:Fig2a-PT-category}.

In the proxy-layer PTs, meek takes the most time of 5.8s to access websites, and snowflake takes the least 2.3s. 
We statistically verify that meek incurs significantly higher time than the rest of the proxy-layer PTs, and snowflake takes significantly lower time than the others (see Appendix Table \ref{tab:Fig2a-second}).
Further, the paired t-test shows a significant difference between meek (M=8.37, SD=4.49) and snowflake (M=3.93, SD=3.72); [t=35.59, \textit{P}<.001]. The 95\% CI is [4.19, 4.68].

\begin{figure*}[h!]
\begin{subfigure}{.45\textwidth}
  \centering
  \includegraphics[scale=0.5]{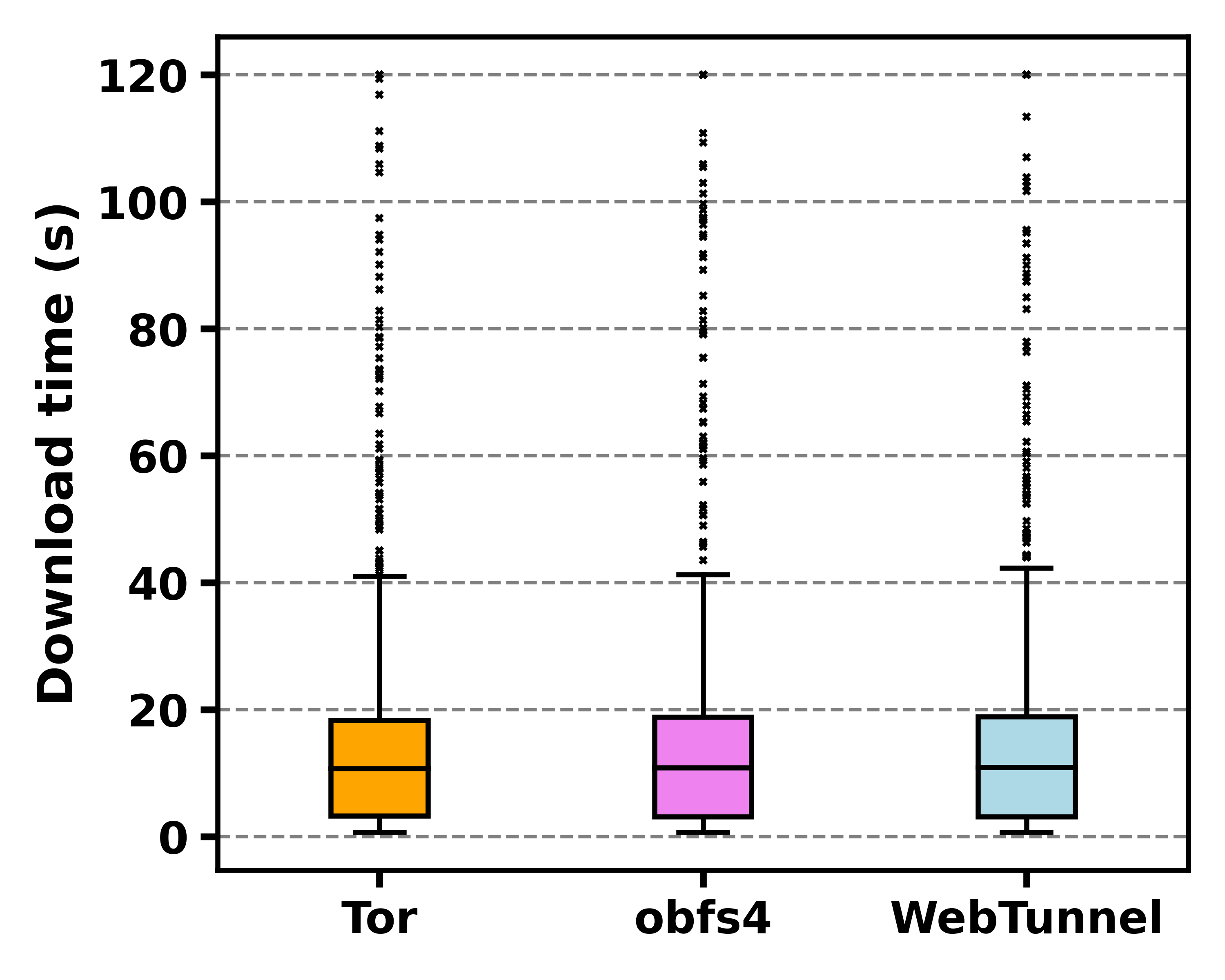}
  \caption{Website access time using vanilla Tor, obfs4, and \\ webtunnel.}  
  \label{fig:web_access_selenium_fix_circuit}
\end{subfigure}
\begin{subfigure}{.45\textwidth}
  \centering
  \includegraphics[scale=0.5]{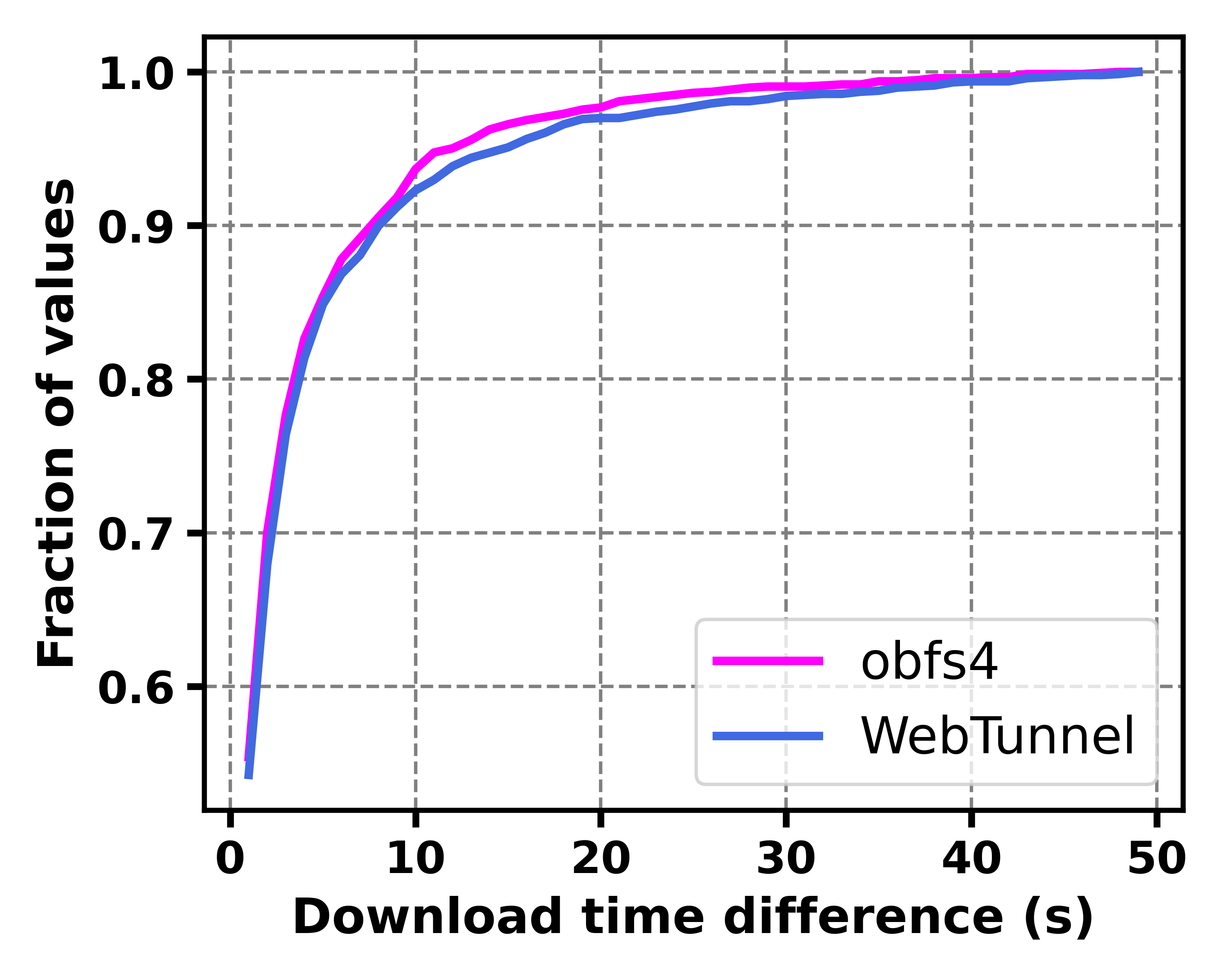}
  \caption{ECDF of time difference for each website via vanilla Tor and PT.}  
  \label{fig:web_access_selenium_fix_circuit_CDF}
\end{subfigure}
\caption{Website access time using a fixed circuit.}
\end{figure*}

Whereas in tunneling-based PTs, camoufler takes the maximum time of 12.8s and webtunnel the lowest with a value of 2.9s. 
We statistically verify the same (see Appendix table \ref{tab:Fig2a-second} for details). The paired t-test shows a significant difference between camoufler (M=16.04, SD=4.74) and webtunnel (M=4.70, SD=3.64);
[t=60.55, \textit{P}<.001]. The 95\% CI is [10.97, 11.70].
Most PTs in the fully encrypted and mimicry category performed well except for marionette, which took an average time of 20.8s. Across all PTs, marionette was the least performing PT, whereas obfs4 and snowflake were the best ones.
The reason for the marionette's low performance can be attributed to the fact that it attempts to obfuscate traffic by programming a user model, which can be harder to optimize for good performance \cite{dyer2015marionette}.
Moreover, the good-performing transports introduce minimal overhead, leading to web access times nearly the same as vanilla Tor.\footnote{The trend for accessing CBL-1k was similar to Tranco top-1k. Thus, we show a common box plot for both sets.} 

For many existing anti-censorship systems, the web access performance is evaluated using the command line utilities (\eg \texttt{curl}) \cite{sharma2020siegebreaker,sharma2021camoufler}. 
However, from the user experience perspective, a better approach would be capturing the actual browsing behavior. When a user types in a URL in the browser, first, the default webpage is downloaded, and subsequently, there are multiple web requests generated by the default webpage to load additional resources (\eg javascript, embedded images). This behavior is not captured by simply using \texttt{curl} to access the default webpage and thus may not reflect the actual performance experienced by a regular user~\cite{ahmad2020apophanies}.
Thus, to observe any potential difference in the download times, we extended our evaluation by accessing the websites using \textit{selenium web browser automation}. Here again, we accessed the Tranco top-1k and CBL-1k websites and recorded the time different PTs took to load all the components of a webpage.
\Cref{fig:web_access_selenium} depicts the average page load time for each website. 
\dg{Here again, we conducted paired t-tests for each pair of PTs and reported the \textit{P}-value, t-value, 95\% CI, and the mean difference in Appendix Tables \ref{tab:Fig2b-first} and \ref{tab:Fig2b-second}.}

As expected, the overall website access time significantly increased for all PTs compared to accessing just the default webpage using \texttt{curl}.
In general, the trend was similar to \texttt{curl} download times, but there were a few deviations. In the proxy-layer PTs, although meek still remained the worst-performing PT, the best-performing PT changed from snowflake to conjure. Snowflake's median time of 32s was almost $2.5\times$ more than conjure (13.7s). 
The paired t-test shows a significant difference between snowflake (M=35.64, SD=22.37) and conjure (M=17.36, SD=14.09);
[t=29.00, \textit{P}<.001]. The 95\% CI is [17.04 to 19.52]. One possible explanation is that the snowflake server was overloaded when we performed selenium-based experiments (see details in \Cref{subsec:snowflake-load-iran}). 
Additionally, we note that in tunneling-based PTs, camoufler could not be evaluated as it does not support multiple simultaneous requests (generated by the selenium browser automation). 

\subsubsection{\textbf{Some PTs performed better than vanilla Tor}}
\label{PTs-better-than-Tor}
PTs modify the Tor traffic, involve additional proxying operations, and some involve additional hop(s) as well. Thus, their download performance should either be inferior (or, in the best case, similar) to Tor. But our experimental results indicate otherwise.
We observed a surprising trend---obfs4, webtunnel, and conjure performed better than vanilla Tor (see \Cref{fig:web_access_selenium}). 
\dg{The paired t-test shows a significant difference between Tor and obfs4 ([t = 16.68,
\textit{P}<.001], the 95\% CI is [5.23, 6.63]), Tor and webtunnel ([t = 8.68, \textit{P}<.001], the 95\% CI is [3.25, 5.14]), Tor and conjure ([t = 7.90, \textit{P}<.001], the 95\% CI is [2.28, 3.79]).}

To investigate this performance difference, we revisited the architectures of these PTs and communicated with the Tor developers who maintain these PTs. 
We realized that Tor manages some high-end servers of these PTs, and there is a consistent effort to optimize for performance.
Since the PT servers (of these PTs) themselves act as the guard relay (see \Cref{subsec:exp-setup}), 
we hypothesized that these PT servers are more optimized for performance compared to a volunteer-operated guard relay. This might explain the potential reason for the observed performance differences.
\vspace{1mm}

\noindent \textbf{Hosting private PT servers:} To confirm the hypothesis, we designed an experiment where rather than using Tor's default PT servers, we deployed our private PT servers.\footnote{We could not host our own conjure server as it needs ISP deployment.} We again accessed these websites involving our PT servers with the expectation that for low-end cloud servers, the download time should be comparable to the vanilla Tor. But, the trend did not change---these PTs incurred lower time to access the websites than direct Tor; for obfs4, the average download time was 19.2s, but for vanilla Tor, it was 24.6s (22\% more than obfs4).
These results challenged our initial hypothesis and indicated that there is something more fundamental than just high-end servers and optimizations that are affecting the performance of these PTs. Thus, we designed a series of experiments to identify the precise reason for such performance differences.\footnote{Initially, we decided to use Ting~\cite{cangialosi2015ting} to identify and quantify the bottleneck in the Tor circuit. But on careful inspection, we found that it cannot be used with PTs (see \Cref{app:ting} for details).}
\vspace{1mm}

\noindent \textbf{Fixing the Tor circuit:} In this set of experiments, we accessed websites using vanilla Tor and PT via a fixed Tor circuit (\ie the same guard, middle, and exit node). 
The rationale for this experiment was to fix the variable components and observe if there is still any difference in performance.

\begin{figure}[h!]
    \centering      \includegraphics[width=0.42\textwidth]{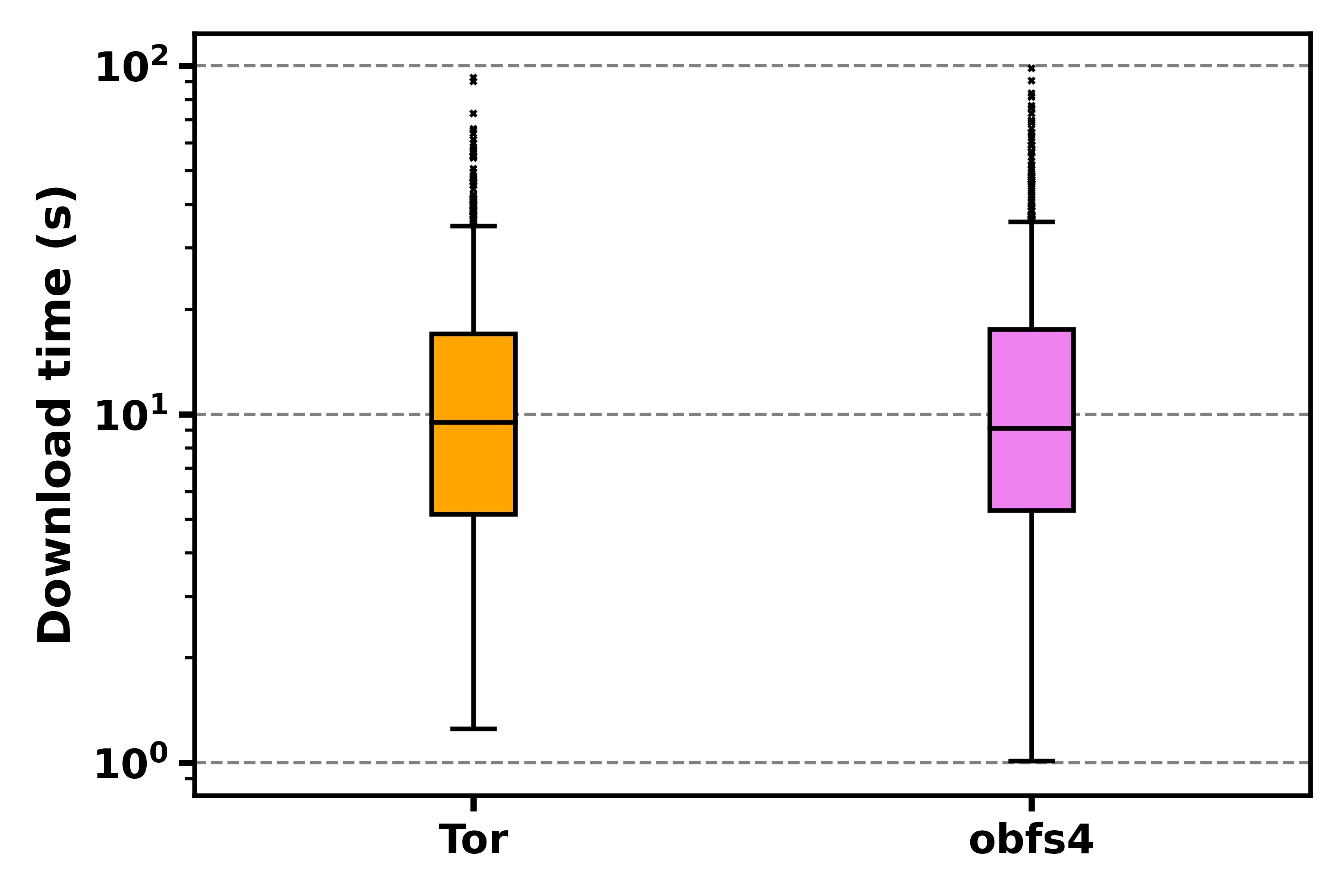}
     \vspace{-2mm}
	\caption{Website access time for Tor and obfs4 using a fixed guard and variable middle and exit nodes. (Y-axis in log scale)}    
 \label{fig:web_access_selenium_fix_guard}
\end{figure}

For both PT and vanilla Tor, fixing the middle and exit nodes is straightforward; however, it's non-trivial to ensure the \textit{same} host as the first hop. This is because vanilla Tor uses a regular guard node, whereas a PT uses the PT server as the first hop of the circuit. Thus, to ensure the same first hop, we configured our guard node and private PT server on the same cloud host. We sampled five Tranco websites (static, news, video streaming, gaming, and online shopping) as target websites for this test. In a single iteration, we accessed each of them using vanilla Tor, obfs4, and webtunnel. Note that in each iteration, we used our own guard (for vanilla Tor) and our private PT server (for obfs4 and webtunnel) as the first hop and a fixed middle and exit node. We performed a total of 500 iterations (for each website), and for every iteration, the middle--exit node pairs were different.

\Cref{fig:web_access_selenium_fix_circuit} shows three box plots of website access time corresponding to vanilla Tor (M=13.41, SD=14.58), obfs4 (M=13.17, SD=\\14.54), and webtunnel (M=13.59, SD=14.81). Ideally, using the same circuit, we should observe similar performance with and without PT for a given website. In \Cref{fig:web_access_selenium_fix_circuit}, we see the expected behavior---\ie nearly identical boxplots. 
Moreover, the results of the paired t-test further indicate that there is a non-significant, very small difference between webtunnel--Tor, obfs4--Tor, and webtunnel--obfs4. For webtunnel--Tor, the 95\% CI is [-0.34, 0.70], with [t=0.66, \textit{P}=0.508]. For obfs4--Tor, the 95\% CI is [-0.72, 0.24], with [t=-0.98, \textit{P}=0.327]. For webtunnel--obfs4, the 95\% CI is [-0.07, 0.91] with [t=1.66, \textit{P}=0.95].

We went a step ahead and made a more nuanced comparison. We analyzed the results for each accessed website individually. To that end, we calculated the absolute time difference of a particular website when accessed via PT and via vanilla Tor.\footnote{Note that these time values are always positive as we apply the modulus function on the difference in time values.}
\Cref{fig:web_access_selenium_fix_circuit_CDF} depicts the ECDF of these time differences. It is evident that for more than 80\% of the cases, the difference in download time was less than 5s. 

Overall, this result establishes that when the relays are the same, the performance of vanilla Tor and PT is also nearly the same. But, during our initial experiments (see \Cref{fig:web_access_selenium}), the circuits and the corresponding relays were not the same. This indicates that the performance difference could be due to the different selected relays.
\vspace{1mm}

\begin{figure}[h!]
	\centering	
    \includegraphics[width=0.44\textwidth]{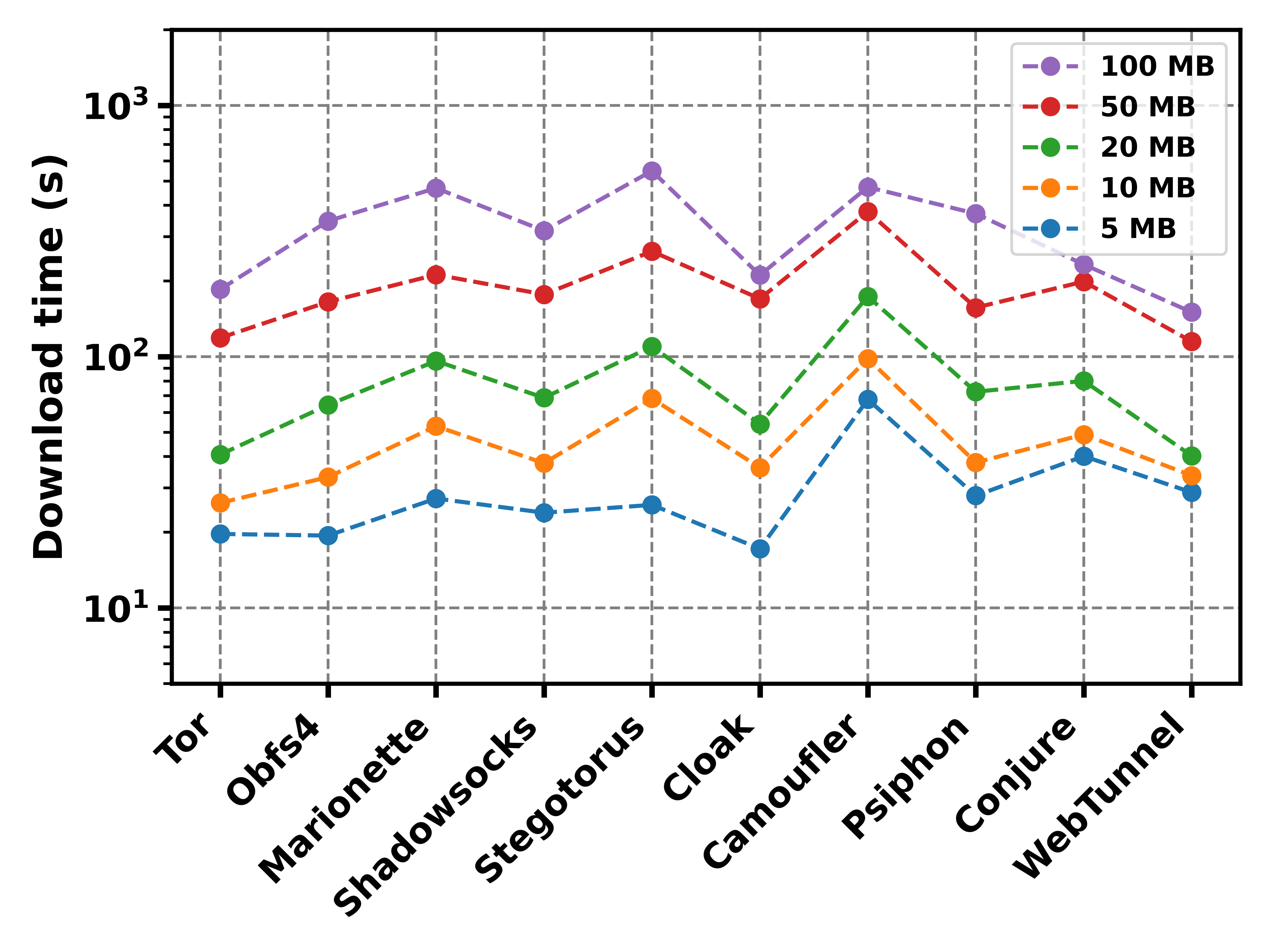}
    \vspace{-2mm}
	\caption{File download time for files of varying sizes using different pluggable transports. (Y-axis in log scale)
 }
	\label{fig:file-download-curl}
\end{figure}

\noindent \textbf{Fixing the guard node:}
We designed our second set of experiments to study the impact of the middle and exit nodes on the observed performance. 
Generally, for a client, the guard node does not change often \cite{guard_spec}.
Thus, in this set of experiments, on the same cloud host, we ran our own guard utility and also the PT server. We used this host as our first hop: for vanilla Tor, we used our guard;  for PTs, we used our private PT server. Middle and exit relays were selected by the Tor client program using Tor's default circuit selection algorithm. We then accessed Tranco top-1k websites via vanilla Tor and the PT.
As can be seen in \Cref{fig:web_access_selenium_fix_guard}, we observed almost the same performance for vanilla Tor and obfs4.\footnote{We repeated the same experiment for the other PT, and we observed a similar trend.}
We repeated the same set of experiments by running our own guard and PT server on the cloud-hosting machines at three geographic locations. The trend was similar; download times of Tranco top-1k were nearly the same.
This strongly suggests that a sufficient variety of middle and exit nodes does not influence the overall performance, and the first hop (guard node/PT server) largely impacts the download performance. 

With this knowledge, we can explain the anomalous behavior of seeing a better performance of some PTs than vanilla Tor. The regular guard nodes are volunteer-run and transfer most of the client traffic seen by the Tor network. But, in contrast, PTs are only used when the default Tor is blocked, and thus, PT servers are generally \textit{less occupied} by clients, leading to better performance. Note that this observation is only for the proxy-layer PTs. Other PTs do not perform better than Tor despite having low client traffic, as the actual bottleneck may be the restrictive nature of the underlying communication method they use.\footnote{Note that there may also be other reasons for this observation (see Sec.~\ref{subsec:snowflake-load-iran} for details).} For instance, dnstt is limited by DNS packet sizes. 

Our experimental results thus indicate that the first hop largely governs the download performance for a Tor circuit. This observation can be extended to the general performance characteristics of the Tor network. However, it is a separate research project in itself, and thus, we keep it out of the scope of current work. 

\begin{figure}[h!]
	\centering	\includegraphics[scale=0.45]{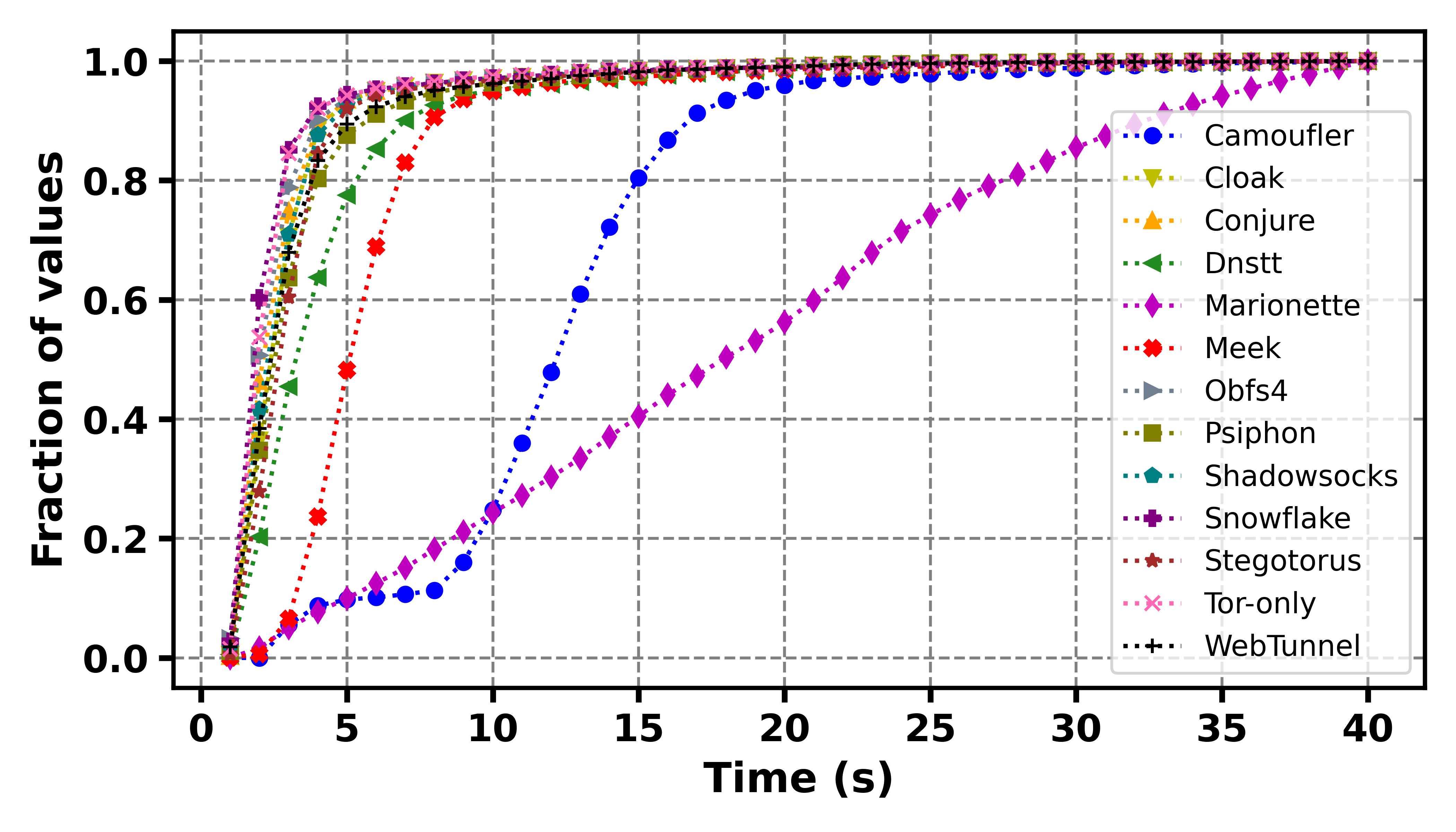}
	\caption{Time to first byte (TTFB) for all pluggable transports when accessing websites.}
	\label{fig:ttfb}
\end{figure}

\subsection{File download time}
\label{eval:file}
This parameter represents the time taken to download files of varying sizes (\ie 5, 10, 20, 50, and 100 MB). It represents the scenario where clients download videos, documents \emph{etc.} via these PTs. To perform the experiments, we set up a cloud server and hosted files of varying sizes on it. We then used different PTs to download each file multiple times (10) via the Tor network. We recorded the average download time (via \texttt{curl}) and plotted it for different PTs in \Cref{fig:file-download-curl}.

In \Cref{fig:file-download-curl}, we see that across file sizes, some PTs (\eg obfs4, cloak) perform well compared to other transports, and some yielded poor performance (\eg marionette, camoufler).
Paired t-test for all PT pairs shows that four PTs, \viz obfs4, cloak, psiphon, and webtunnel, performed significantly better than the remaining PTs, and there were no significant differences among these four PTs.
In contrast, these tests also reveal that marionette's download time is significantly more than that of other PTs. See Appendix \Cref{tab:fig5-first} for details.

For a file size of 10 MB, obfs4 and cloak took, on average, 33s and 36s. Other transports took considerably more time, with camoufler taking 98s, almost $3\times$ more than obfs4. 
The high download time for camoufler can be attributed to the rate limit imposed by IM providers to send and receive content via their APIs. We obtained similar results for selenium-based file downloads, with obfs4, cloak, and conjure performing better than other transports.

Note that there were other PTs that did not succeed in downloading a file at all. 
In \Cref{fig:file-download-curl}, we only show the values for PTs that succeeded in downloading files of each size at least twice. Dnstt, snowflake, and meek could not do that for most of the file sizes in our experiments and thus are excluded from the figure. 
For instance, meek could download files of different sizes only once out of ten attempts, with significantly high download times. For a 5 MB file, it required 110.5s; for 10 MB, it took 224.9s; for 50 MB, it incurred 1028.8s; a 100 MB file was downloaded in 1558.9s.
We quantify this unreliable behavior of the PTs failing to download content in detail in \Cref{eval:reliability}.  
Overall, our results suggest that
PTs such as obfs4, cloak, psiphon, and webtunnel can be used for fast downloads, whereas others can be used to achieve satisfactory performance.

\begin{figure}[h!]
	\centering
	\includegraphics[scale=0.47]{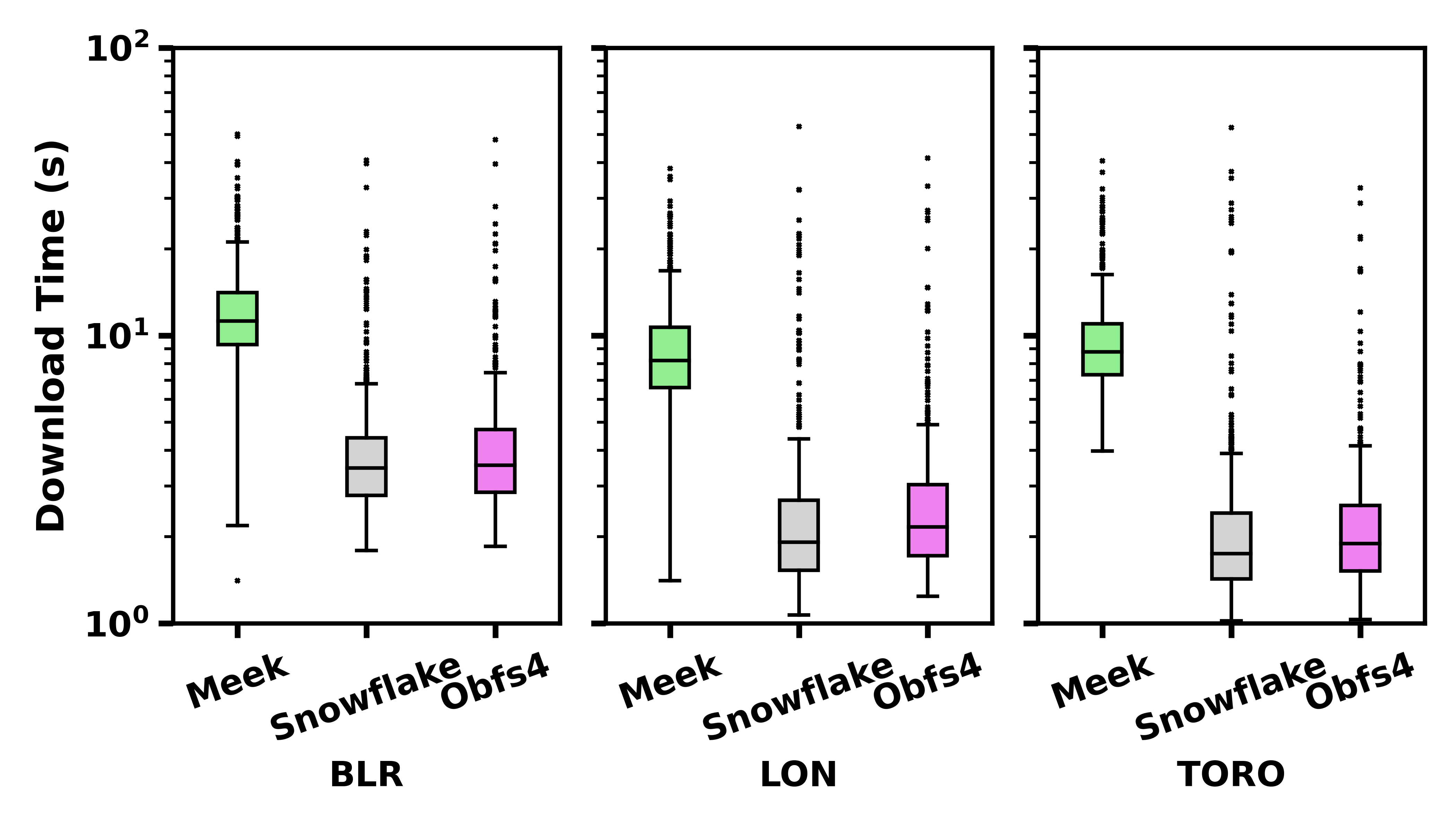}
	\caption{Website access time for meek, obfs4, and snowflake for different locations: Bangalore (BLR), London (LON), Toronto (TORO). (Y-axis in log scale)}
	\label{fig:location}
\end{figure}

\begin{figure*}[h!]
      \begin{subfigure}{0.6\textwidth}
      \centering
     \includegraphics[scale=0.6]{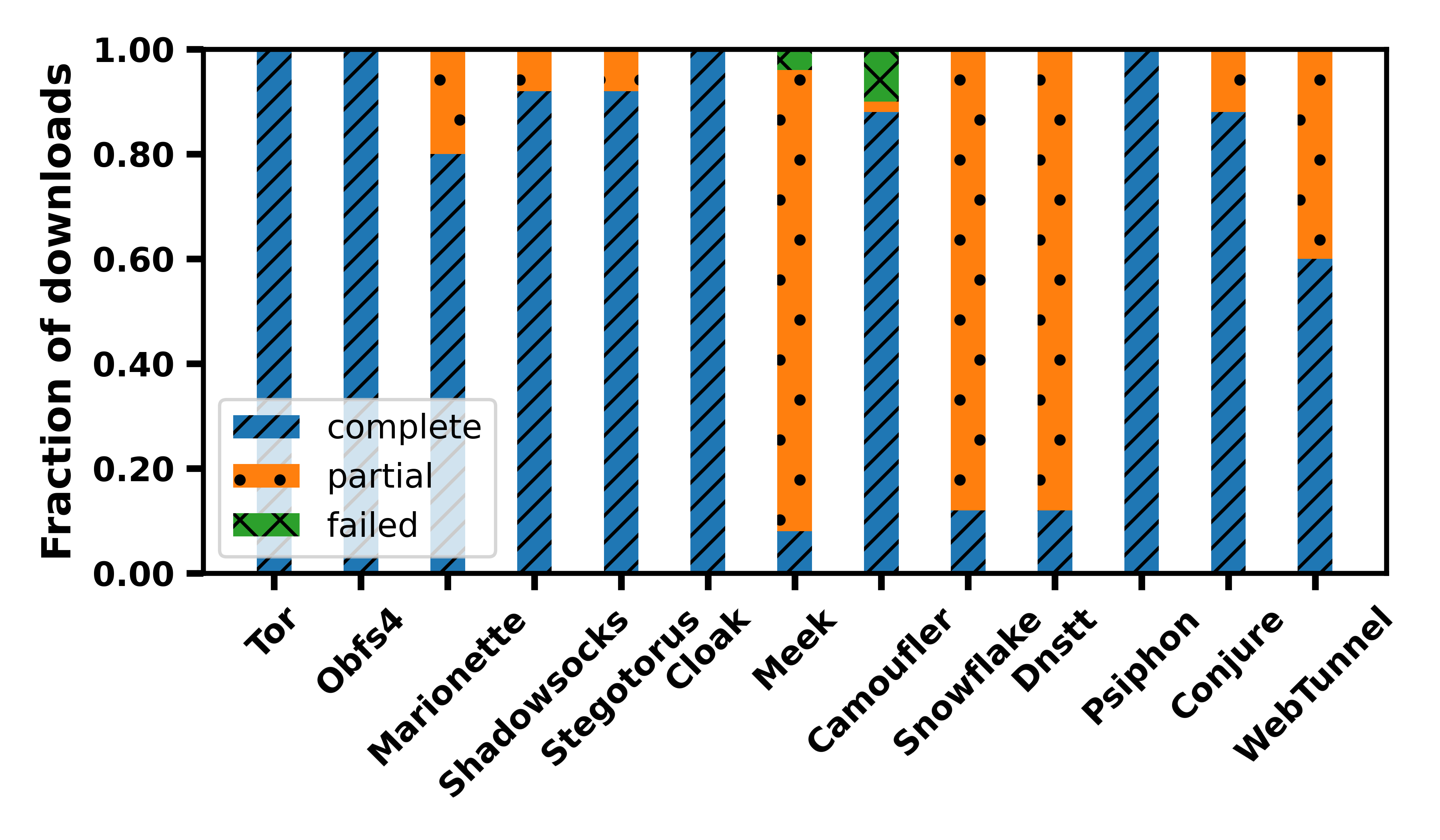}
	\caption{Fraction of complete, partial, or failed file downloads.}
    \label{fig:reliability}
     \end{subfigure}
    \begin{subfigure}{0.37\textwidth}
    \centering
     \includegraphics[scale=0.5]{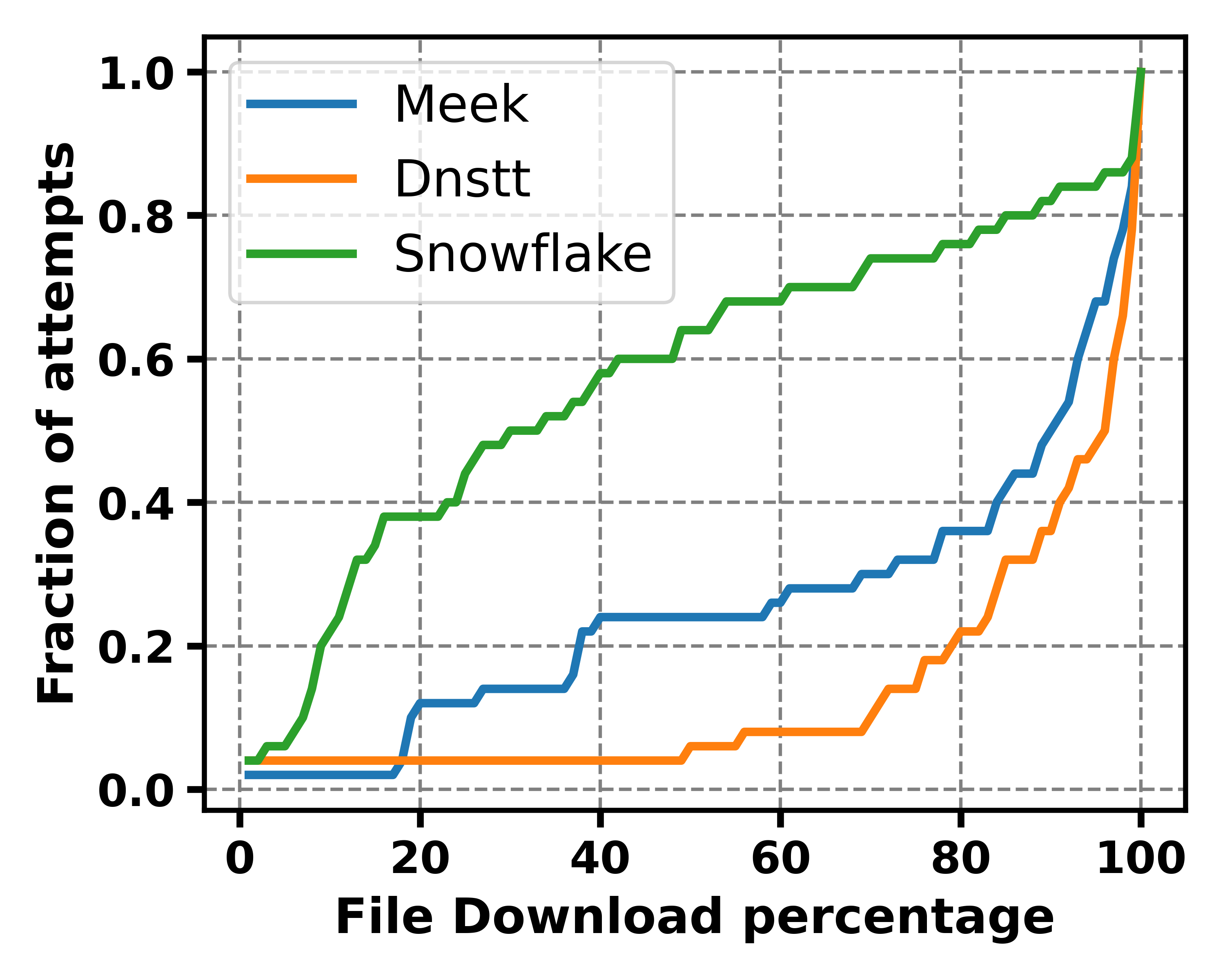}
	\caption{ECDF of fraction of file download attempts vs. portion of files downloaded.}   
        \label{fig:file-partial-download}
     \end{subfigure}
     \caption{Reliability of pluggable transports.}
\end{figure*}

\subsection{Time to first byte (TTFB)} 
TTFB is the time difference between initiating a request for a web resource and receiving the first byte of application data. Thus, it captures the initial bootstrap latency incurred by the PT. If the TTFB is high, the transport may not be well suited for interactive applications such as web browsing. 

We show the ECDF of TTFB values for all PTs across all websites in \Cref{fig:ttfb}. Except for meek (shown as \color{red} \textbf{x} \color{black}), marionette (shown as \color{violet}$\smallblackdiamond$\color{black}), and camoufler (shown as \color{blue}$\smallblackcircle$\color{black}), the response time of other PTs is relatively small, with more than 80\% of the websites getting their first data byte in less than 5s. Marionette has the highest response time, with about 40\% of websites taking more than 20s to get their first byte. In meek, the response time is between 2.5s to 7.5s for about 90\% of the websites, whereas in camoufler, it was 2.5s to 17.5s for the same.

The potential reason for meek and camoufler resulting in high response time is likely due to their design considerations. For instance, meek involves substantial initial processing. The fronting service (to which the client connects) completes the TLS handshake (with the meek client), decrypts the HTTPS request, and then forwards it to the actual intended server. Also, the meek bridge is rate-limited by its maintainer \cite{meek-throttle}. This could also be the reason for high TTFB.
Thus, any PT except for meek, marionette, and camoufler can be considered for applications that require low response time.

\subsection{Impact of location} 
Clients residing at different geographical locations might experience different performances for the same PT. Thus, we measure the performance of a PT by varying client and server locations among a total of six countries across three continents. We selected two locations each in North America (New York and Toronto), Europe (Frankfurt and London), and Asia (Bangalore and Singapore). We then considered three locations for clients (Bangalore, London, and Toronto) and three for servers (Singapore, Frankfurt, and New York) and performed the experiments for all 9 (3x3) possible client-server combinations. 

Here again, we accessed websites and downloaded files via all the PTs.
As an example, in \Cref{fig:location}, we show the website access time of meek, snowflake, and obfs4 across the three different client locations. 
It can be seen that the \textit{trend} of snowflake and obfs4 being better than meek remains consistent across different locations. For other PTs also, we observed a similar performance trend as the one shown in \Cref{fig:web_access}.
Moreover, irrespective of the server location, we observed that the access time was higher when the client was in Bangalore than in London or Toronto. This is because most Tor relays are hosted in Europe and North America \cite{tor-relay-loc}, and thus, clients' traffic from Asia may have to travel longer geographical distances. We obtained a similar performance trend when downloading files of different sizes.

\subsection{Reliability}
\label{eval:reliability}
As previously mentioned in \Cref{eval:file}, some PTs may not always be able to download the complete content. We characterize this behavior by collecting instances where a particular file download (or website access) was \textit{not at all} completed or \textit{partially} completed.  
We count such instances for all PTs across all experiments.

In \Cref{fig:reliability}, we show these values as a stacked bar plot for file downloads. 
It is clear that dnstt, meek, and snowflake do not perform well while downloading files. In more than 80\% of the instances, the files were only \textit{partially} downloaded by these. Moreover, in camoufler and meek, in about 10\% of the instances, the file was \textit{not at all} downloaded. 

In \Cref{fig:file-partial-download}, we plot the ECDF of the portion of the file downloaded in different attempts. The figure contains the result for meek, dnstt, and snowflake as we observed maximum unsuccessful download attempts with these PTs. We downloaded different file sizes (5, 10, 20, 50, and 100 MB) 20 times each and recorded what fraction of a file was downloaded to the client machine. We can see that in 60\% of the download attempts, snowflake downloaded less than 40\% of any file. But, meek and dnstt were able to download more content, \ie less than 92\% and 96\% of the total file size. In just about 10\%--20\% of total attempts, we observed a complete download.

The reason meek cannot download files reliably is likely because the public meek bridge is rate-limited \cite{meek-throttle}.\footnote{We contacted the developers of the meek bridge, and they confirmed the same.} Since file access requires downloading significantly more data than accessing websites, the impact of rate limiting seems more pronounced.

In dnstt, data transmission is limited by the underlying DNS packet sizes (see \Cref{sec:background}), and snowflake servers observed an unprecedented increase in users from Iran (see \Cref{subsec:snowflake-load-iran}).
As snowflake downloaded the least content for most of the incomplete file downloads, we may infer that an increase in user load at the PT server more drastically impacts the overall performance than other causes of performance degradation (\eg constraints imposed by the underlying communication primitives).
But there could be other potential reasons for such frequent incomplete file downloads, \eg multiple proxy transitions in an ongoing snowflake connection. If the snowflake proxy changes while downloading a file, it may lead to failure.  
In the future, we plan to conduct a detailed investigation to identify the precise reason for these failures.

Although for website access, meek, dnstt, and snowflake resulted in higher web access times than other PTs (see \Cref{fig:web_access_selenium}), we did not observe this unreliable behavior of incomplete webpage download by them. Hence, they can still be used for accessing websites. A potential reason for this behavior is that file download requires maintaining a connection for a long time compared to web access. Thus, there is a higher chance of connection termination if the PTs have high resource utilization. The remaining PTs, however, can be reliably used for both website access and file downloads. 

Moreover, such unreliable behavior of the PTs may falsely lead the user to believe that the PT is blocked, but in practice, it is just not able to perform well. This can be detrimental to the reputation of the PT and beneficial for the censor as the PT might not be accessed by the clients without the censor having to block it. 

\subsection{Effect of transmission medium}

All our previous experiments were performed with client machines connected via Ethernet. Thus, we conducted experiments to study the impact of change in connection medium on PT performance. We configured the Tor client and PT client utility on lab machines. The machines were connected to the Internet with a more error-prone wireless medium (WiFi). While conducting the experiments, we ensured that our lab WiFi routers were not congested. We accessed the Tranco top-500 and CBL-500 websites (five times each), using all aforementioned PTs, and recorded the download times. 
Overall, we did not see any observable change in the trends when we switched the medium compared to wired connections (see \Cref{eval:web}).
For example, on average, meek incurred 16.4s, whereas dnstt, cloak, and obfs4 took 5.1s, 3.9s, and 3.7s time to access the websites.

Note that in all our experiments, we simply changed the medium (from wired to wireless) and did not introduce any congestion at the router (\eg by introducing more clients or artificially deteriorating the signal strength). 
Thus, a detailed study may be performed to analyze the PT ecosystem in the wireless medium under different settings. We consider such a study out of scope for the current project and keep it as a part of our future work. 
\section{Discussion}

\subsection{Ethical Considerations}
\label{sec:ethics}
In this research, we conducted a large-scale measurement campaign involving the live Tor network and Tor-supported pluggable transport infrastructure. Tor network is used by millions of users, and thus, we ensured that our measurements should not impede regular Tor users. We followed the principles prescribed in the Belmont report~\cite{beauchamp2008belmont}. The three important principles are:
\begin{itemize}
    \item \textbf{Respect for persons:} We did not involve human subjects in our studies and thus did not require obtaining consent.
    \item \textbf{Beneficence:} In our study, we evaluate the performance of PTs over Tor, which can ultimately result in enhanced user experience, offering benefits to the users of the system.
    \item \textbf{Justice:} No user apart from the authors was involved in the experiments; thus, this study does not pose risks to any third-party users. On the contrary, the outcomes of our research can benefit users for whom censorship is a daily reality.  
\end{itemize}

Since our experiments involved sending traffic over Tor, we carefully planned our measurements (spread out across multiple weeks) to not overload the Tor infrastructure. We conducted our experiments from VPSes hosted on cloud infrastructure at different locations and did not involve any residential networks.
We use standard Tor client utility similar to any regular Tor user. Throughout our experiments, we only accessed Tranco top-1k websites and CBL-1k, and file sizes of no more than 100 MB (hosted on our servers only). We took extra caution while performing our experiments involving pluggable transports that are used \textit{en masse} \eg Instant Messaging apps (camoufler), DNS resolvers (dnstt) \etc We ensured that at any point in time, we do not burden these systems (by running experiments in small batches).

For some experiments, we hosted our own Tor guard nodes on cloud-hosting machines. We used them only for accessing websites and downloading file sizes necessary for experiments. We never recorded any personally identifiable information of users (\eg IP address) connecting our guard nodes. Moreover, we hosted them only for the duration of the experiment. Across all our experiments, we hosted a guard node maximum for a duration of four weeks after it received the guard status.

\begin{figure}
	\centering
    \includegraphics[width=0.4\textwidth]{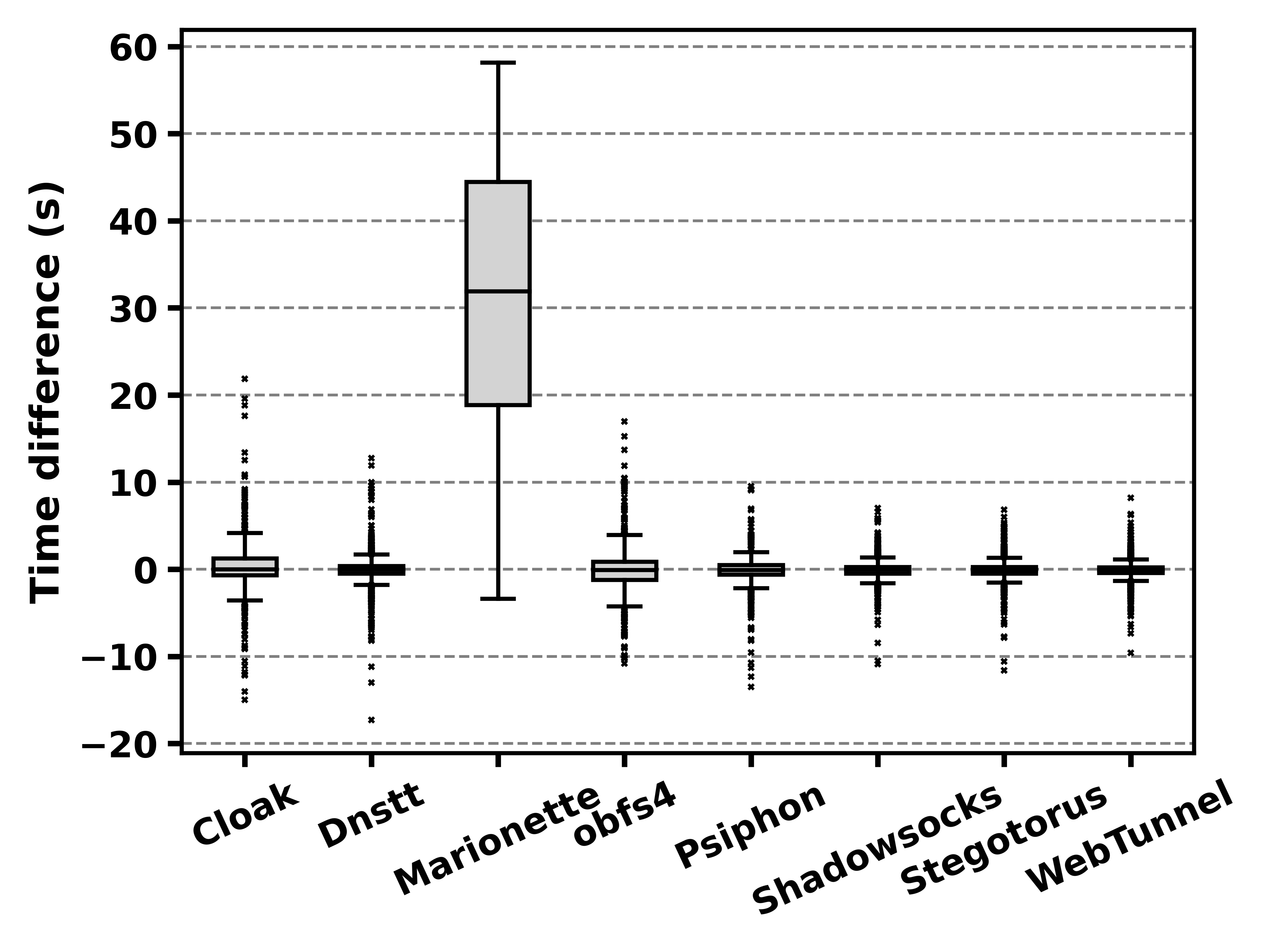}
    \vspace{-3mm}
	\caption{Time difference between pluggable transports and vanilla Tor: A positive value implies PT incurred more time than vanilla Tor, and a negative value implies vice-versa.}
	\label{fig:ptdiff}
\end{figure}

\begin{figure*}[h!]
\centering
\begin{subfigure}{.45\textwidth}
	\centering
	\includegraphics[scale=0.43]{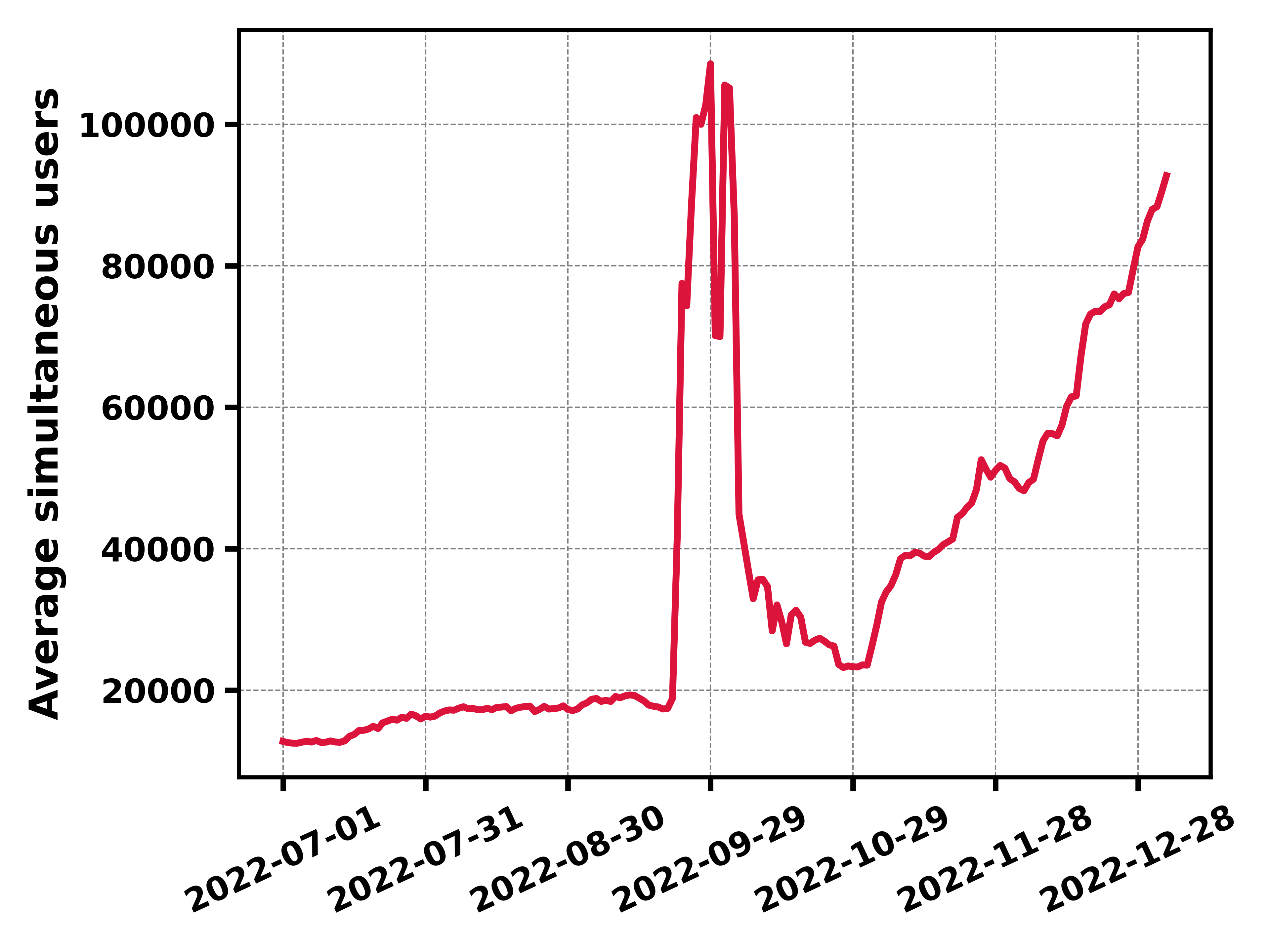}
	\caption{Number of Snowflake users \cite{snowflake_graphs}.}
	\label{fig:Snowflake-users}
\end{subfigure}
\begin{subfigure}{.45\textwidth}
	\centering
	\includegraphics[scale=0.4]{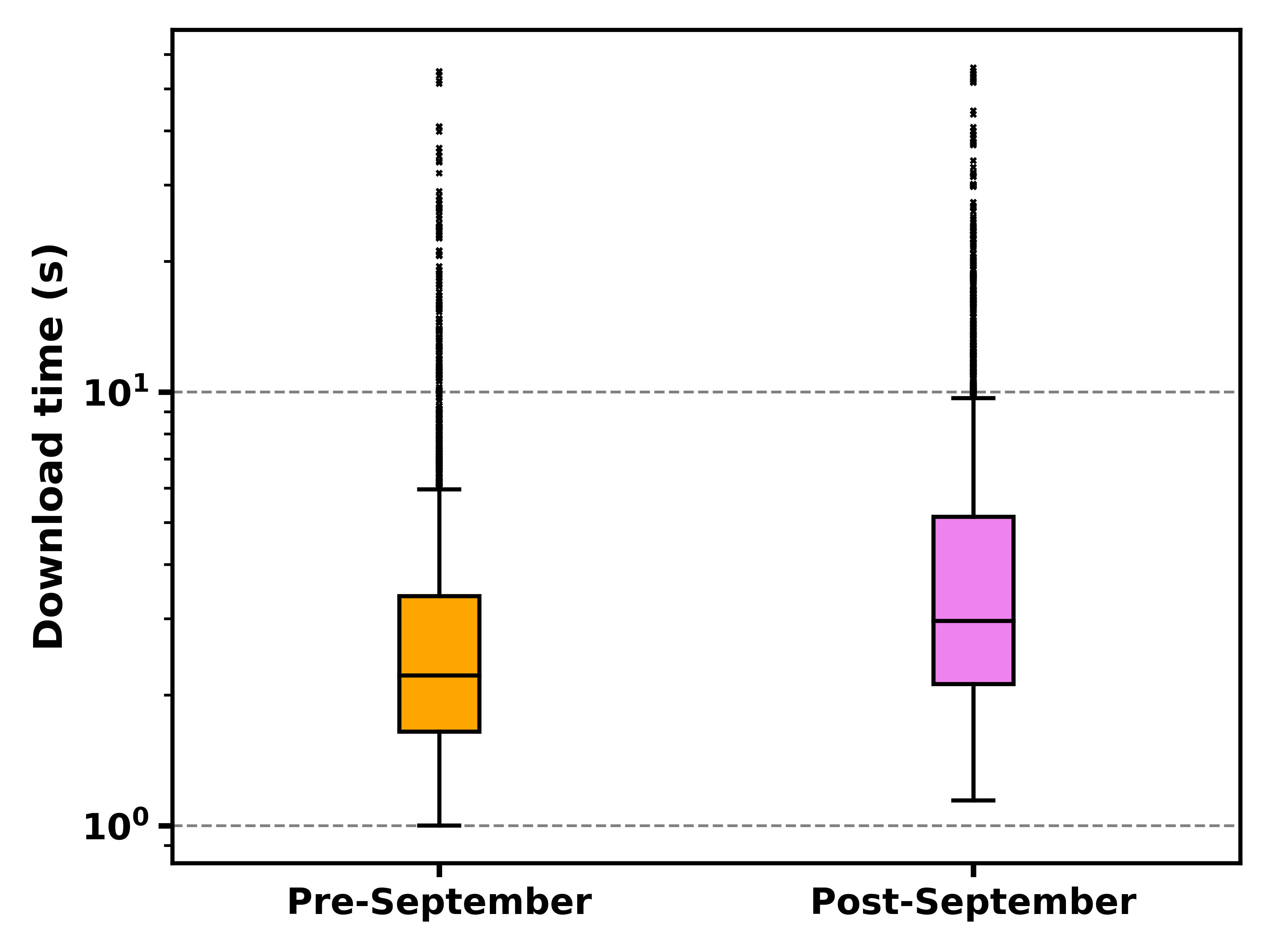}
	\caption{Snowflake performance before and after Iran protests. (Y-axis in log scale.)}
	\label{fig:snowflake-iran-results}
 \end{subfigure}
 \caption{The impact of increasing load on snowflake's performance.}
\end{figure*}

\subsection{Evaluation of PTs without Tor}
\label{subsec:pts_without_tor}
Our PT measurements in earlier sections were conducted over the Tor network and are thus representative of how the PTs perform when they are used in conjunction with Tor. 
However, it is useful to measure the performance overhead solely due to the PTs (\ie without the involvement of Tor). PTs that result in low-performance overhead can be important to other overlay networks (\eg Nym \cite{diaz2021nym}) that desire censorship resistance \cite{Nym-censorship} without relying on third-party anti-censorship solutions (\eg VPN, Tor). These networks can integrate good-performing PTs directly with their network infrastructures. 

Thus, we analyzed if it's possible to \textit{isolate} the effect of PTs on the observed performance (\ie the download time). It would be very convenient to perform such an evaluation if all the PTs had the capability to run independently (without Tor). In such a case, we would download websites (1) via the PTs alone and (2) download the same websites directly over the Internet. The difference between the two would estimate the performance overhead by the PT (if any).
But this is not the case; a good fraction of these PTs (5 out of 12) by default are built to work only in conjunction with the Tor network and not independently. These include obfs4, webtunnel, meek, snowflake, and conjure.

Thus, we used an alternate strategy to measure the performance of all 12 PTs (irrespective of whether PTs can work with/without Tor). The key idea is to access websites first via PT+Tor and then only via vanilla Tor. The difference between the two would estimate the PT overhead (if any).
We now explain how we designed the experiments for PTs that cannot be isolated from Tor and the ones that can be isolated.
\vspace{1mm}

\noindent \textbf{PTs inseparable from Tor:} As previously mentioned, for such PTs, the PT server and the first hop in the Tor circuit are the same. 
But for measuring the PT overhead, we require the client to access a website through a fixed circuit (guard, middle, and exit)---via (1) the PT+Tor and (2) vanilla Tor. Thus, we needed to ensure that each node in the circuit was the same for vanilla Tor and PT. To make the first hop in the circuit identical, we deployed the PT server and ran our own guard node on the same cloud host. Subsequently, we configured the Tor client utility to use the same middle and last hop for the remaining part of the Tor circuit.

We then recorded the time it took to access the website via vanilla Tor and the PTs. The difference in the time provided us with the overhead (if any) caused by the PT for that particular website. We performed this evaluation for Tranco top-1k websites. Each website was accessed using a different circuit, but as described earlier, for a particular website, the circuit remained the same for Tor as well as the PTs. 
Even with this approach, we could not evaluate all those PTs that are inseparable from Tor. For some PTs, controlling the PT server is relatively difficult due to deployment hurdles such as hosting proxy on the CDN in meek, ISP collaboration in conjure \etc Hence, we analyzed three (obfs4, dnstt, and webtunnel) out of the six PTs with this capability.
\vspace{2mm}

\noindent \textbf{PTs separable from Tor:}
We similarly designed experiments for PTs that allow the PT server to run independently of the Tor network.
In this case, selecting the same Tor circuit for the PT and vanilla Tor was relatively easier than in the previous case. This is because the PT server and guard node were separate, and we can specify the complete circuit from the Tor client utility itself. 
Notably, we wanted to capture the overhead of these PTs \textit{only} due to their underlying technology. This required us to minimize the impact of external factors, such as the delay due to the packets traveling from the PT client to the PT server. To that end, we deployed the PT client and server in the same cloud location.

We plot the difference in download times for the PTs for which we could perform this evaluation in \Cref{fig:ptdiff}. It can be seen that most of the PTs did not introduce any significant overhead due to their functioning. This can be attributed to the fact that most of these PTs introduce an extra layer of a proxy, and we minimize the impact of the added latency due to one hop by keeping the PT client and server in the same location. The only exception is marionette, for which we could quantify its overhead with our approach (average website access time is more than 30s). Marionette's main aim is to obfuscate the user's traffic with some cover traffic, such that the two seem indistinguishable from each other. To do so, it uses probabilistic automata at its core, and each transition between the states has an associated action (\eg encrypting a message).

\subsection{Increased user load on snowflake}
\label{subsec:snowflake-load-iran}

In September 2022, massive protests and civil unrest erupted inside Iran \cite{IR-protest-UN, IR-protest-NL}. As a consequence, the government of Iran imposed severe restrictions on Internet access (including blocking the Tor network \cite{IR-block-Tor}). Thus, Iranian citizens resorted to using snowflake pluggable transport to access the Tor network \cite{IR-snow-users}, and the number of users grew abruptly in the last week of September. \Cref{fig:Snowflake-users} shows that after September 2022, a large number of users were actively using snowflake. In October 2022, users decreased drastically \cite{snowflake-issue} due to the blocking of snowflake using TLS fingerprint~\cite{snowflake_tls}.
In November 2022, the issue was resolved by the snowflake maintainers \cite{snowflake_tls}. Since then, we have seen an overall increasing trend in users connecting to the snowflake servers.

Interestingly, our results correlate with these observations. Our pre-September 2022 snowflake PT experiments yield comparatively better performance than post-September 2022 experiments. Using \texttt{curl}, we accessed the Tranco top-1k websites pre-September 2022 and post-September 2022. 
In \Cref{fig:snowflake-iran-results}, we see that post-September 2022\footnote{We performed this experiment in November as the snowflake server was not stable in October.}, the average web access time has increased.
The paired t-test also shows a significant difference between pre-September (M=3.42, SD=4.30) and post-September (M=4.77, SD=5.42);
[t=-10.76, \textit{P}<.001]. The 95\% CI is [-1.59, -1.10].

Moreover, we attempted to download a small file of size 5 MB, but post-September 2022, in 8 out of 10 attempts, we failed. We repeated this experiment five times once every week. In all attempts, in the majority of the cases, we were unable to download the complete file.
This further supports our observation that the PT server (or guard node) largely impacts the download performance (see \Cref{PTs-better-than-Tor}).

It also explains the anomalous behavior of snowflake with selenium (see \Cref{fig:web_access_selenium}); the website load time is much higher compared to the snowflake with \texttt{curl} experiments \ie more than five times (see \Cref{fig:web_access}). The increase is not only due to the use of selenium but also due to the fact that we performed the selenium-based experiments starting in November 2022 with a high user load.\footnote{Very recently, Fifiled and Nordberg \cite{fifieldrunning} also report the problem of high traffic load on snowflake bridges and even proposed solutions to manage the increasing load.}

However, since we observed that the user load on snowflake servers increased (in September end), we performed our post-September measurements with extra caution. We did not want to add undue load on the already overloaded server. Thus, we performed only 100--200 measurements in a day. This led us to complete the post-September measurements in months. We further continued performing the experiments for the subsequent months to monitor the change in performance with fluctuations in the number of users. However, the number of snowflake users did not decrease after the substantial increase in September 2022, and thus, the observed average download time remained consistently greater than what was seen pre-September (see \Cref{app:snowflake_cont}
for details).

\subsection{Performance evaluation  using speed index}
\label{app:speedindex}

\begin{figure}
	\centering	\includegraphics[width=\columnwidth]{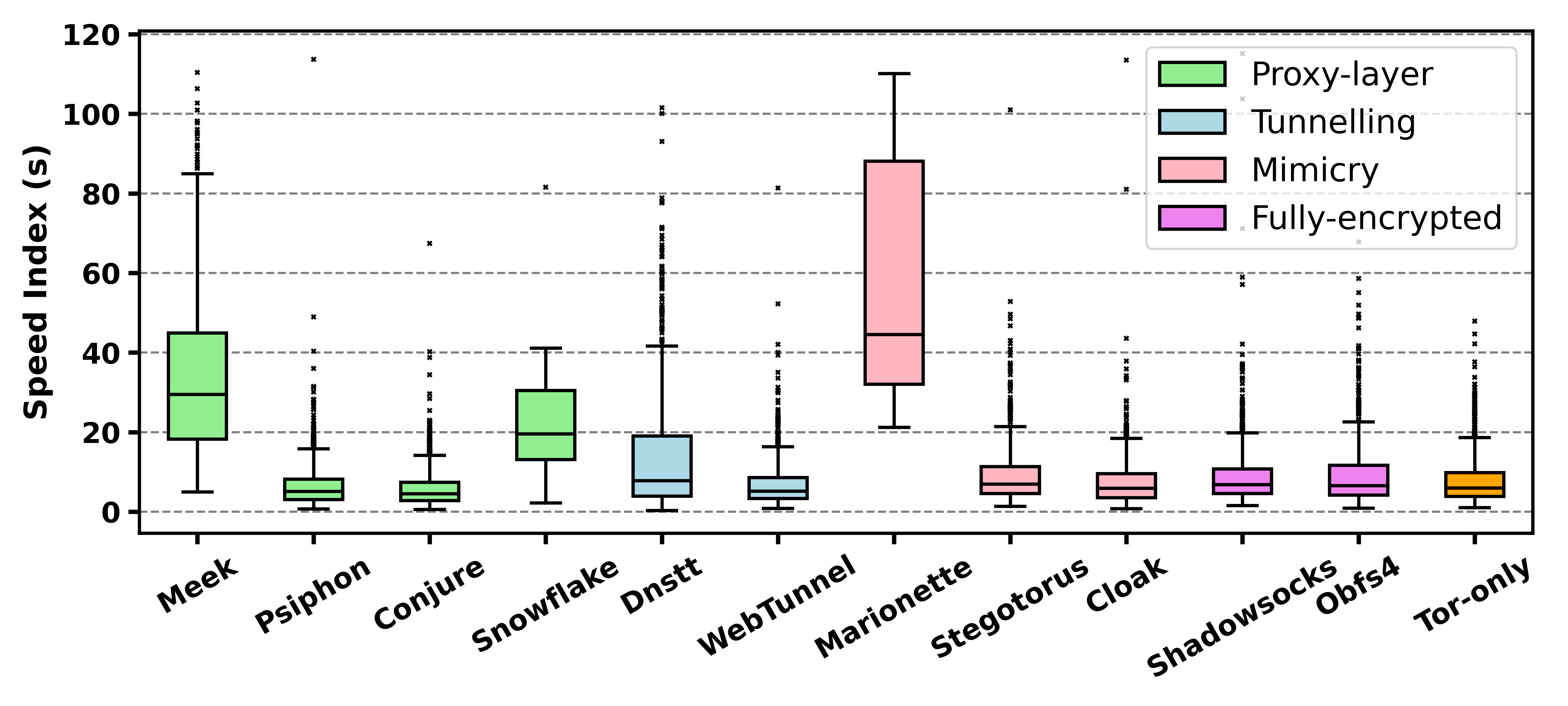}
	\caption{Speed index for all pluggable transports using browsertime.}
	\label{fig:speed_index}
\end{figure}

Our evaluation of PTs involves calculating the website access time, TTFB \etc, using curl and selenium-based automation.
However, recently other performance metrics such as speed index~\cite{speedIndex} have also been proposed. The speed index captures the time it takes to load all the visual elements of a webpage and provides a nuanced perspective on usability. Thus, we performed additional experiments to calculate the speed index for the PTs (using browsertime framework~\cite{browsertime}). \Cref{fig:speed_index} depicts the results. 
We observe that the trend among individual PT categories and across the categories remains consistent with our results from the selenium-based evaluation. 
\dg{For instance, in proxy-layer based PTs, meek incurs the maximum time, and in mimicry-based PTs, marionette takes the highest time.
The results of the paired t-test between all combinations of PT pairs are summarized in Appendix Tables \ref{tab:Fig11-first} and \ref{tab:Fig11-second}.}

Additionally, one can see that the speed index is lower for all PTs, signifying that the users will be able to visualize the webpage much before all the elements of the webpage are loaded. However, note that this lowering of the time is exclusively the property of the webpage itself, and the PTs themselves do not optimize for identifying and loading the visual elements.

\subsection{Limitations}
\label{subsec:limitations}

Our research provides valuable insights into the PT ecosystem. However, it is important to consider certain limitations when interpreting our results. 

First, our measurements are time-gapped, and this might impact the trends and statistical significance of the results. Since Tor is a volunteer-operated network (often with limited bandwidth relays), we intentionally time-gapped the measurements so as not to overload this network (following the best ethical practices).

Second, there are confounding factors that can also impact the download time. For instance, the traffic load on relay nodes, background Internet traffic, \etc are some of the external factors \cite{DDoS_Tor,Tor-DDoS-seven} that can further impact our measurements. Thus, in some of our experiments, we even used our own hosted Tor relays to reduce the impact of external traffic on the relays (see \cref{PTs-better-than-Tor}). Moreover, as a general rule, we conducted multiple measurements at different times on the actual Tor network to minimize the impact of such factors.

Third, the PT performance in censored countries might deviate from what we report in the paper. Although we consider multiple vantage points to study the impact of location, we could not get access to vantage points hosted within censored countries (\eg Iran, Russia, and China). The sophisticated censorship infrastructure deployed within these countries might further affect the performance 
of the PTs \cite{palfrey2007}.

Overall, while our study provides useful insights, real-world complexities introduce some limitations. We took reasonable steps to minimize variability, but some factors remain beyond our control.
\section{Conclusion}
Pluggable Transports (PTs) for Tor are becoming increasingly important to stay ahead in the censorship arms race. Thus, it is essential to evaluate different aspects of PTs (\eg unobservability, performance, usability) for their continual improvement.
In this paper, we conduct a first performance evaluation of all Pluggable Transports (PTs) presently used in Tor and those that can be integrated with Tor in the future. Out of the 28 PTs that we analyzed, we were able to run and test 12 PTs. Amongst the remaining 16 PTs, 13 are non-functional; two are for specific use cases (\eg messaging), and one has restricted access (requires a passcode from developers).

For the 12 functional PTs, we recorded the website access time, file download time, time to first-byte \etc, from different locations around the globe. 
Our results show that the PT performance is largely impacted by the underlying technology (\eg content tunneled inside DNS packets) and load at the PT server. Moreover, not all PTs can be used to access different types of content. 
For instance, meek and dnstt cannot completely download a file most ($80\%$) of the time. These frequent failed download attempts could be inimical to PT's reputation as clients may falsely believe that PTs are subjected to blocking.

Overall, our study highlights that crucial aspects like PTs' performance warrant attention from the research community. Users need to be made aware of the right choice of PT, depending upon the application they would use. Otherwise, it may result in the inaccessibility of the content, which may cause user fatigue.\\

\noindent \textbf{Acknowledgments:} We thank our shepherd, Nguyen Phong Hoang, and the anonymous reviewers for their thoughtful and encouraging inputs in the entire reviewing process. We also thank Tarun Kumar Yadav for having insightful discussions about how to apply statistical tests and deduce meaningful inferences correctly. We express
our gratitude towards David Fifield and Cecylia Bocovich for providing us with their detailed constructive feedback on the initial draft of this paper. We also thank Sambuddho for graciously providing us with unhindered infrastructural support (\ie cloud hosting machines) to conduct these measurements. Moreover, this research is partially supported by the Research Council KU Leuven under the grant C24/18/049, by CyberSecurity Research Flanders with reference number VR20192203, and by DARPA FA8750-19-C-0502. Any opinions, findings, conclusions, or recommendations expressed in this material are those of the authors and do not necessarily reflect the views of any of the funding agencies.

\bibliographystyle{ACM-Reference-Format}
\bibliography{reference}

\balance

\appendix

\section{Appendix}

\subsection{Pluggable Transports at a glance (a comparative analysis)}
\label{comparison}

\begin{table*}
\centering
\begin{tabular}{|p{2.5cm}|p{1.35cm}|p{1.4cm}|p{1.6cm}|p{1.7cm}|p{3.4cm}|p{3.1cm}|}
 \hline
 \multicolumn{7}{|c|}{\textbf{PTs bundled in the Tor Browser}} \\
 \hline
 \textbf{Name} &   \textbf{Code available} &   \textbf{Functional} &   \textbf{Integratable} &   \textbf{Performance evaluated} &   \textbf{Implementation challenges} &  \textbf{ Underlying technology}\\
 \hline
  Obfs4 \cite{obfs4} & \cmark & \cmark & \cmark & \cmark  &   None &   Random obfuscation \\
 \hline
 Meek \cite{meek} &  \cmark & \cmark & \cmark & \cmark  &   Requires CDN with domain fronting support &   Domain fronting \\
 \hline
  Snowflake \cite{snowflake} & \cmark & \cmark & \cmark & \cmark  &   Dependency on domain fronting &  WebRTC \\
 \hline
 
 \multicolumn{7}{|c|}{\textbf{PTs listed by the Tor project and currently under deployment/testing}} \\
 \hline

  Dnstt \cite{dnstt} & \cmark & \cmark & \cmark & \cmark &   None &   DoH/DoT tunneling  \\
 \hline
  Conjure \cite{torConjure}  & \cmark & \cmark & \cmark & \cmark 
  &   Needs ISP support &   Decoy routing \\
 \hline
WebTunnel \cite{webTunnel} & \cmark & \cmark & \cmark & \cmark &   None &   Tunneling over HTTP  \\
 \hline
TorCloak \cite{torCloakmail, torCloakpdf} & \xmark & N/A & N/A & N/A & N/A &   Tunneling over WebRTC \\
 \hline
 
 \multicolumn{7}{|c|}{\textbf{PTs listed by the Tor project but undeployed}} \\
 \hline

 Marionette \cite{dyer2015marionette} & \cmark & \cmark & \cmark & \cmark  &   Dependency issues (supports only Python 2.7) &   Network traffic obfuscation \\
 \hline
 Shadowsocks \cite{shadowsocks} & \cmark & \cmark & \cmark & \cmark  &   None &   Network traffic obfuscation \\
 \hline
 Stegotorus \cite{weinberg2012stegotorus} & \cmark & \cmark & \cmark & \cmark  &   None &   Steganographic obfuscation\\
 \hline
  Psiphon \cite{psiphon} & \cmark & \cmark & \cmark & \cmark  &   None &   Proxy-based\\
 \hline
 Lantern Lampshade \cite{lanternlampshade} & \cmark & \xmark & \xmark & N/A &   Unavailability of ready to deploy code  &   Obfuscated encryption\\
 \hline
 
 \multicolumn{7}{|c|}{\textbf{PTs neither listed nor deployed by the Tor Project}} \\
 \hline

 Cloak \cite{cloak} & \cmark & \cmark & \cmark & \cmark &   None &   Network traffic obfuscation \\
 \hline
 Camoufler \cite{sharma2021camoufler} & \cmark & \cmark & \cmark & \cmark  &   Dependency on IM    accounts &   Tunneling over IM application\\
 \hline
 Massbrowser \cite{nasr2020massbrowser} & \cmark & \cmark & \cmark & \cmark(partial)  &   Requires invite-code from authors &   Domain fronting and browser based proxy \\
 \hline
 Protozoa \cite{protozoa} & \cmark & \xmark & \xmark & \xmark &   Code compilation issues &   Tunneling over WebRTC \\
 \hline
 Stegozoa \cite{figueira2022stegozoa} & \cmark & \xmark & \xmark & \xmark & Provides basic functionality, sends only text data over sockets &   Tunneling over WebRTC \\
 \hline
 Sweet \cite{zhou2012sweet} & \cmark & \xmark & \xmark & N/A &   Dependency issues &   Tunneling over emails\\
 \hline
 DeltaShaper \cite{barradas2017deltashaper} & \cmark & \xmark & \xmark & N/A &   Requires Skype version that is no longer supported &   Tunneling over video\\
 \hline
 Rook \cite{vines2015rook} & \cmark & \cmark & \xmark & N/A &   Can only be used for messaging; no proxy support &   Hiding data using online gaming \\
 \hline
 Facet \cite{facet} & \cmark & \xmark & \xmark & N/A &   Requires Skype version no that is longer supported  &   Tunneling over video \\
 \hline
 Mailet \cite{mailet} & \cmark & \cmark & \xmark & N/A &   Can only be used to access Twitter; no proxy support  &   Tunneling over email \\
 \hline
 MinecruftPT \cite{minecruft} & \cmark & \xmark & \xmark & N/A &  Issues in the source code \cite{issueMine} &   Hiding data using online gaming \\
 \hline
 CloudTransport \cite{brubaker2014cloudtransport} & \xmark & N/A & N/A & N/A & N/A &   Tunneling over cloud\\
 \hline
 CovertCast \cite{mcpherson2016covertcast} & \xmark & N/A & N/A & N/A  & N/A &   Tunneling over video\\
 \hline
 FreeWave \cite{houmansadr2013want} & \xmark & N/A & N/A & N/A & N/A &   Tunneling over VoIP \\
 \hline
 Balboa \cite{rosen2021balboa} & \xmark & N/A & N/A & N/A & N/A &   Obfuscation based on user-traffic model\\
 \hline
 Domain Shadowing \cite{wei2021domain} & \xmark & N/A & N/A & N/A & N/A &   Domain shadowing \\
 \hline
\end{tabular}

\caption{Comparison of Pluggable Transports}
\label{table:allpt}
\end{table*}

In \Cref{sec:background}, we provided a brief description of different PTs that were evaluated in our study. However, 
the regular attempts to block Tor by China \cite{dunna2018analyzing}, coupled with the recent events of excessive banning of Tor and PTs by different regimes \cite{IR-block-Tor, Russia-ban-net3}, highlights the pressing need for more feasible options for PTs. Thus, in this work, we studied a total of 28 systems that could become PTs. But, many were not part of our experiments for various reasons. With the help of Table~\ref{table:allpt}, we provide an overview of all 28 PTs and the challenges involved in their adoption. 

The parameters in the table are suitably chosen for comparison. For instance, in the first row, for obfs4, we mention that its code was available, it is functional and integrated into Tor, and thus we measured its performance.
But for torCloak the code is not publicly available, and thus we can not test it to ascertain whether it can be integrated with Tor. Thus, we mark not applicable (\texttt{N/A}) in the respective fields.
Additionally, we could only partially test the working of the massbrowser. This is because it requires an access code to function for each unique device. We obtained only one access code from the authors, and thus, we could test it from a single vantage point. For a fair comparison, we do not include it in our analysis, as we tested other PTs from several geographic locations.
Similarly, we contacted the developers of all those PTs whose source codes were available, but we encountered errors while installing and running them. We succeeded in running and testing some PTs (\eg webtunnel) based on the response from their developers. However, for some PTs, it was not possible to fix the bugs with the suggested changes.


Next, depending upon their adoption status by the Tor project, we categorize these PTs broadly into four categories---(1) PTs bundled with the Tor browser, (2) PTs listed by the Tor project and are under deployment and testing, (3) PTs listed by the Tor project and are not under deployment (4) PTs not listed by Tor.
Currently, only three PTs (obfs4, meek, and snowflake) are integrated with the Tor browser and can be easily used by the clients. 
Four are under deployment testing (\ie dnstt, conjure, webtunnel, and torcloak). The remaining 21 PTs are not under consideration for adoption. 
As previously stated, we aimed to deploy and run all these systems for performance comparisons. 
However, we found that 13 of the considered PTs were non-functional. We found multiple issues with them \eg source code not available, compilation errors, and deprecated versions of the underlying technology (mentioned in the table). Our analysis reveals that more focus should be given to the reproducibility of the proposed systems. 
Thus, we make our code and analysis scripts public at \cite{ptperf_github} to foster future research on PTs.

\subsection{Snowflake post-September monitoring}
\label{app:snowflake_cont}

\begin{figure}[h]
    \includegraphics[width=0.4\textwidth]{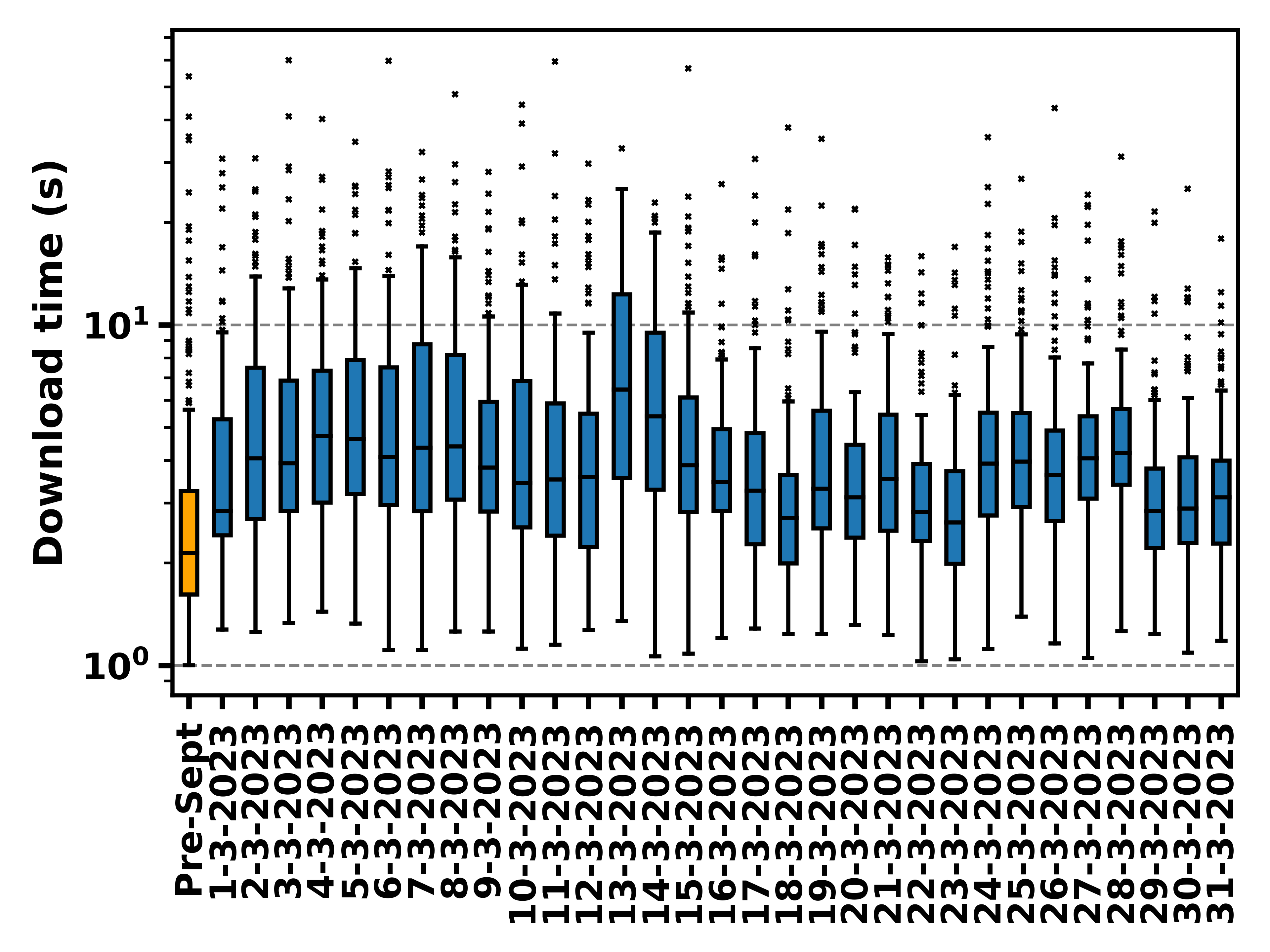}
	\caption{Snowflake performance before and after Iran protests. Note that the y-axis is in the log scale.}
	\label{fig:snowflake_cont}
\end{figure}

We measured the performance of snowflake for March 2023 after the unrest in Iran (refer \Cref{subsec:snowflake-load-iran}). 
We randomly selected 100 websites from Tranco top-1k websites and downloaded each website five times using \texttt{curl}. \Cref{fig:snowflake_cont} first shows the box plot of measurements conducted before the unrest (in orange color). Second, all subsequent box plots (in blue color) correspond to measurements conducted in March 2023 (after the unrest). We can observe that after the unrest, the average download time is always high compared to pre-September download values.

\subsection{Implementation Details}
\label{sec:implemnt}
We deployed all PT clients and server machines on the Digital Ocean cloud hosting service. We used Ubuntu 20.04 (kernel v5.15 generic) on all machines.

\textbf{PT specific implementation details:}
For snowflake, meek, obfs4, and conjure, we do not host our own PT server instances. Instead, we directly use the ones provided by the Tor Project page \cite{torPT} 
due to deployment hurdles in hosting them. For instance, meek requires the server to be hosted on a CDN with domain fronting support involving monthly subscription fees; conjure requires placement of a server within an ISP (see \Cref{table:allpt} for more details). This can act as a deterrent for the end users to host their own servers. Thus, wherever possible, we use the PT servers provided by the Tor project, as our goal is to study PT performance perceived by the users.
For the rest of the PTs, we hosted our own PT server instances in different geographical locations. 

Furthermore, dnstt requires a domain name to be registered (see \Cref{sec:background}). Thus, we registered a domain name on Namecheap hosting service \cite{nameC} and added multiple sub-domains to it. All these sub-domains further pointed towards custom authoritative nameservers, which are actually PT server machines corresponding to each sub-domain. 

\textbf{Selenium setup:} For chrome browser automation, we selected \texttt{Selenium-v4.4.3} along with \texttt{Chromedriver-v105.0.5195.52} (as a webdriver). Chrome web driver provides flexible and custom capabilities to add a wide range of proxy protocols over different ports. To block advertisements on websites fetched by selenium, we added \texttt{uBlock origin} \cite{uBlock} as an extension to Chrome. We added appropriate options such as \texttt{--no-sandbox, --headless} in the selenium browser instance and set a page load timeout of 120 seconds with each instance for website download. In the case of file downloads, we set the default minimum timeout to 1200s so that selenium can take enough time to download files of size 100MB (maximum file size in our experiments).\footnote{For PTs that resulted in incomplete file downloads with a 1200s timeout value (see \Cref{fig:reliability}), we increased their timeout to 7200s to provide them additional time to download the content. However, the results did not change.} We used the same timeout for our \texttt{curl} based file download experiments.

\textbf{Browsertime setup}: To measure the speed index, we used browsertime \cite{browsertime}---a framework to perform web-related measurements and record values of various performance metrics. It takes multiple arguments as input parameters such as chrome options, URLs to observe, number of iterations, and type of metrics (\eg visual metrics like speed index). For the experiments, we specify the same chrome options as in the previous Selenium setup (\eg chrome driver version \texttt{105.0.5195.52}, page load timeout of 120s \etc). Upon completion of the specified iterations, browsertime generates a JSON file specifying the value of the speed index (in ms) that we use for plotting results.

\textbf{Tor circuit selection:} Some of our experiments in \Cref{eval:web} required fixing the complete circuit or just the guard (middle or exit) nodes in a circuit. To fix the nodes, the underlying PT code communicates with Tor over the control port specified in the \texttt{torrc} file. 
In our scripts, we used \texttt{stem library} \cite{stemLib} to instruct Tor not to make new circuits on their own. 
Thus, we set \texttt{MaxClientCircuitsPending} configuration parameter to 1. 
Also, we ensured that for the duration of our experiment, the created circuit should persist.  
Hence, we also set parameters \texttt{NewCircuitPeriod} and \texttt{MaxCircuitDirtiness} to sufficiently high integer values. 
The higher the value, the more the retention time of the circuit. 
Once these parameters are set, Tor does not create new circuits on its own.
Further, in \Cref{subsec:pts_without_tor}, we set the value of \texttt{LeaveStreamsUnattached} to 1 and use \texttt{carml} \cite{carml} to instruct Tor to attach incoming streams to our created specific circuits.

\subsection{Future Directions}
In our study, we considered the most common use cases for accessing a website or downloading files. 
Thus, in the future, other use cases, \eg audio streaming, could be explored for evaluating PTs' performance.
We note that our analysis can be extended to more websites, client locations (in censored countries) \emph{etc.} We envision that periodic performance measurements of deployed PTs could also be integrated with the Tor project for long-term analysis.

Additionally, apart from performance, there were other usability concerns that we encountered while evaluating the PTs. For instance, camoufler requires creating an account on the IM app, which in turn involves a mobile number for registration. This may lead to usability challenges. Similarly,
massbrowser's performance can also be evaluated, but it requires (per device) access code from the authors. 
Thus, in the future, a detailed usability study of PTs can also be conducted such that a wider audience can use them.

\subsection{On using Ting to identify the bottleneck in Tor circuit}
\label{app:ting}

In \Cref{PTs-better-than-Tor}, we report that obfs4 and webtunnel performed better than vanilla Tor. We subsequently conducted a series of experiments to understand why that is the case. All such experiments involved creating Tor circuits with different combinations of fixed and variable relays to identify the bottlenecks in performance. However, it could be argued that one could use Ting~\cite{cangialosi2015ting} (an approach for measuring latency between arbitrary Tor relays) to identify the bottlenecks. Ting can be used to quantify the delay introduced between the PT server/guard node and the middle node and compare it with the delay between the middle and the exit node.

However, it is not possible to use Ting to measure the latency when PTs are involved. This is because Ting requires setting up a circuit in which the first hop in the circuit should be controlled by the Ting operators while the second hop should be the Tor node that one wants to measure the latency about.
\footnote{Note that there are a total of three different configurations of circuits in Ting. In all the configurations, the first and the last hop Tor nodes should be controlled by Ting operators, while the nodes in between them are the ones for which the latency is to be measured. These nodes can be any of the guard, middle, or exit nodes. 
}
For circuits involving PTs, the PT server can \textit{only} act as the first hop, and \textit{cannot} be used as a second hop in the circuit. 
Thus, PT-based circuits do not satisfy the necessary conditions required by Ting. Hence, Ting cannot be used to measure the latency for PT-based circuits.

Note that we also considered modifying Ting for measuring circuit delays involving PTs. However, any modifications in Ting (specific to PTs) violated its assumptions.
For instance, with one of our modifications, we could have approximated the latency between the PT server and the middle node. However, this involves an assumption that the regular TCP packets and Tor packets are \textit{not} treated differently by the ISPs. This assumption was shown in Ting to yield potentially inaccurate results; thus, we did not consider using it.



\begin{table*}[htbp]
\centering
\begin{tabular}{cccccc}
\toprule
PT Pair               & CI Lower & CI Upper & t-value & \textit{P}-value         & Mean diff. \\ \midrule
Tor-Dnstt             & -5.222   & -4.359   & -21.751 & \textless{}.001 & -4.791     \\
Cloak-Dnstt           & -5.541   & -4.736   & -25.028 & \textless{}.001 & -5.138     \\
Stegotorus-Dnstt      & -4.948   & -4.140   & -22.040 & \textless{}.001 & -4.544     \\
Marionette-Camoufler   & 2.229    & 4.035    & 6.798   & \textless{}.001 & 3.132      \\
Dnstt-Psiphon         & 3.824    & 4.684    & 19.401  & \textless{}.001 & 4.254      \\
Obfs4-Conjure         & -1.328   & -0.828   & -8.450  & \textless{}.001 & -1.078     \\
Stegotorus-WebTunnel  & -0.935   & -0.348   & -4.281  & \textless{}.001 & -0.641     \\
Snowflake-Camoufler   & -12.793  & -12.040  & -64.553 & \textless{}.001 & -12.416    \\
Stegotorus-Snowflake  & 0.425    & 0.850    & 5.876   & \textless{}.001 & 0.638      \\
Shadowsocks-Snowflake & -0.566   & -0.069   & -2.506  & 0.01            & -0.318     \\
Shadowsocks-WebTunnel & -1.929   & -1.263   & -9.393  & \textless{}.001 & -1.596     \\
Tor-Cloak             & 0.129    & 0.659    & 2.918   & 0.004           & 0.394      \\
Tor-Psiphon           & -0.721   & -0.164   & -3.117  & 0.002           & -0.443     \\
Tor-Obfs4             & 0.824    & 1.441    & 7.196   & \textless{}.001 & 1.133      \\
Snowflake-WebTunnel   & -1.528   & -0.933   & -8.094  & \textless{}.001 & -1.230     \\
Shadowsocks-Meek      & -4.972   & -4.426   & -33.750 & \textless{}.001 & -4.699     \\
Tor-Conjure           & -0.184   & 0.244    & 0.273   & 0.78            & 0.030      \\
Cloak-Camoufler       & -12.895  & -12.173  & -68.022 & \textless{}.001 & -12.534    \\
Cloak-Conjure         & -0.506   & -0.104   & -2.974  & 0.003           & -0.305     \\
Dnstt-WebTunnel       & 3.549    & 4.421    & 17.908  & \textless{}.001 & 3.985      \\
Tor-Stegotorus        & -0.428   & 0.074    & -1.383  & 0.17            & -0.177     \\
Meek-Psiphon          & 3.328    & 3.854    & 26.761  & \textless{}.001 & 3.591      \\
Stegotorus-Cloak      & 0.380    & 0.740    & 6.097   & \textless{}.001 & 0.560      \\
Tor-Meek              & -4.305   & -3.882   & -37.896 & \textless{}.001 & -4.094     \\
Obfs4-Snowflake       & -1.012   & -0.492   & -5.672  & \textless{}.001 & -0.752     \\
Tor-WebTunnel         & -1.110   & -0.491   & -5.067  & \textless{}.001 & -0.800     \\
Obfs4-Shadowsocks     & -0.597   & -0.329   & -6.774  & \textless{}.001 & -0.463     \\
Obfs4-Camoufler       & -13.561  & -12.854  & -73.149 & \textless{}.001 & -13.208    \\
Obfs4-Meek            & -5.407   & -4.826   & -34.479 & \textless{}.001 & -5.117     \\
Psiphon-Conjure       & 0.291    & 0.740    & 4.501   & \textless{}.001 & 0.516      \\
Cloak-Meek            & -4.633   & -4.118   & -33.277 & \textless{}.001 & -4.375     \\
Camoufler-Conjure     & 11.789   & 12.479   & 68.891  & \textless{}.001 & 12.134     \\
Marionette-Shadowsocks & 15.038   & 16.393   & 45.453  & \textless{}.001 & 15.715     \\
Marionette-Conjure     & 14.424   & 15.730   & 45.255  & \textless{}.001 & 15.077     \\
Tor-Snowflake         & 0.086    & 0.608    & 2.606   & 0.009           & 0.347      \\
Obfs4-Stegotorus      & -1.496   & -1.085   & -12.328 & \textless{}.001 & -1.291     \\
Meek-Camoufler        & -8.216   & -7.490   & -42.424 & \textless{}.001 & -7.853     \\
Snowflake-Conjure     & -0.526   & -0.142   & -3.408  & \textless{}.001 & -0.334     \\
Obfs4-Cloak           & -0.968   & -0.605   & -8.495  & \textless{}.001 & -0.786     \\
Marionette-Stegotorus  & 14.210   & 15.522   & 44.413  & \textless{}.001 & 14.866     \\
Shadowsocks-Camoufler & -13.165  & -12.441  & -69.359 & \textless{}.001 & -12.803  \\
\bottomrule
\end{tabular}
\vspace{2mm}
\caption{Paired t-test results for PTs in \Cref{fig:web_access}---website access using \textbf{curl} [Part I].}
\label{tab:Fig2a-first}
\end{table*}

\begin{table*}[htbp]
\centering
\begin{tabular}{cccccc}
\toprule
PT Pair                & CI Lower & CI Upper & t-value & \textit{P}-value         & Mean diff. \\ \midrule
Snowflake-Dnstt        & -5.594   & -4.763   & -24.439 & \textless{}.001 & -5.178     \\
Obfs4-Dnstt            & -6.370   & -5.530   & -27.779 & \textless{}.001 & -5.950     \\
Tor-Camoufler          & -12.417  & -11.648  & -61.336 & \textless{}.001 & -12.032    \\
Cloak-Psiphon          & -1.121   & -0.564   & -5.933  & \textless{}.001 & -0.843     \\
Marionette-Snowflake    & 14.764   & 16.047   & 47.071  & \textless{}.001 & 15.406     \\
Camoufler-Dnstt        & 6.961    & 7.953    & 29.467  & \textless{}.001 & 7.457      \\
Obfs4-Psiphon          & -1.907   & -1.332   & -11.039 & \textless{}.001 & -1.620     \\
Psiphon-WebTunnel      & -0.662   & -0.052   & -2.293  & 0.02            & -0.357     \\
Shadowsocks-Dnstt      & -5.966   & -5.118   & -25.625 & \textless{}.001 & -5.542     \\
Marionette-Dnstt        & 9.426    & 11.038   & 24.879  & \textless{}.001 & 10.232     \\
Tor-Shadowsocks        & 0.455    & 1.023    & 5.105   & \textless{}.001 & 0.739      \\
Meek-Dnstt             & -1.183   & -0.354   & -3.634  & \textless{}.001 & -0.768     \\
Camoufler-WebTunnel    & 10.974   & 11.708   & 60.555  & \textless{}.001 & 11.341     \\
Meek-Conjure           & 3.878    & 4.327    & 35.859  & \textless{}.001 & 4.102      \\
Tor-Marionette          & -15.717  & -14.442  & -46.393 & \textless{}.001 & -15.079    \\
Camoufler-Psiphon      & 11.145   & 11.978   & 54.428  & \textless{}.001 & 11.561     \\
Marionette-Meek         & 10.313   & 11.628   & 32.717  & \textless{}.001 & 10.970     \\
Stegotorus-Camoufler   & -12.238  & -11.590  & -72.067 & \textless{}.001 & -11.914    \\
Cloak-Snowflake        & -0.208   & 0.275    & 0.272   & 0.79            & 0.033      \\
Dnstt-Conjure          & 4.425    & 5.241    & 23.208  & \textless{}.001 & 4.833      \\
Marionette-Cloak        & 14.708   & 16.043   & 45.159  & \textless{}.001 & 15.376     \\
Shadowsocks-Cloak      & -0.524   & -0.174   & -3.906  & \textless{}.001 & -0.349     \\
Marionette-WebTunnel    & 13.482   & 14.824   & 41.342  & \textless{}.001 & 14.153     \\
Shadowsocks-Psiphon    & -1.419   & -0.880   & -8.358  & \textless{}.001 & -1.150     \\
Cloak-WebTunnel        & -1.487   & -0.894   & -7.873  & \textless{}.001 & -1.190     \\
Stegotorus-Psiphon     & -0.566   & -0.109   & -2.896  & 0.004           & -0.338     \\
Meek-WebTunnel         & 2.959    & 3.565    & 21.080  & \textless{}.001 & 3.262      \\
Stegotorus-Meek        & -4.112   & -3.612   & -30.310 & \textless{}.001 & -3.862     \\
Marionette-Psiphon      & 13.888   & 15.200   & 43.461  & \textless{}.001 & 14.544     \\
Shadowsocks-Stegotorus & -1.051   & -0.674   & -8.955  & \textless{}.001 & -0.863     \\
Snowflake-Psiphon      & -1.060   & -0.623   & -7.545  & \textless{}.001 & -0.842     \\
Conjure-WebTunnel      & -1.185   & -0.614   & -6.179  & \textless{}.001 & -0.899     \\
Obfs4-Marionette        & -16.843  & -15.478  & -46.418 & \textless{}.001 & -16.161    \\
Stegotorus-Conjure     & 0.100    & 0.472    & 3.019   & 0.003           & 0.286      \\
Obfs4-WebTunnel        & -2.408   & -1.712   & -11.607 & \textless{}.001 & -2.060     \\
Snowflake-Meek         & -4.685   & -4.196   & -35.599 & \textless{}.001 & -4.440     \\
Shadowsocks-Conjure    & -0.873   & -0.426   & -5.706  & \textless{}.001 & -0.649     \\
\bottomrule
\end{tabular}
\vspace{2mm}
\caption{Paired t-test results for PTs in \Cref{fig:web_access}---website access using \textbf{curl} [Part II].}
\label{tab:Fig2a-second}
\end{table*}

\begin{table*}[htbp]
\centering
\begin{tabular}{cccccc}
\toprule
PT Pair               & CI Lower & CI Upper & t-value & \textit{P}-value         & Mean Diff. \\ \midrule
Tor-Dnstt             & -21.627  & -18.545  & -25.540 & \textless{}.001 & -20.086    \\
Cloak-Dnstt           & -21.733  & -18.716  & -26.271 & \textless{}.001 & -20.224    \\
Stegotorus-Dnstt      & -14.508  & -8.500   & -7.505  & \textless{}.001 & -11.504    \\
Dnstt-Psiphon         & 20.319   & 24.575   & 20.679  & \textless{}.001 & 22.447     \\
Obfs4-Conjure         & -3.417   & -1.967   & -7.279  & \textless{}.001 & -2.692     \\
Stegotorus-WebTunnel  & 10.456   & 16.249   & 9.036   & \textless{}.001 & 13.352     \\
Stegotorus-Snowflake  & -11.294  & -4.189   & -4.271  & \textless{}.001 & -7.742     \\
Shadowsocks-Snowflake & -10.525  & -8.253   & -16.200 & \textless{}.001 & -9.389     \\
Shadowsocks-WebTunnel & 9.108    & 11.454   & 17.175  & \textless{}.001 & 10.281     \\
Tor-Cloak             & -0.621   & 0.872    & 0.329   & 0.74            & 0.125      \\
Tor-Psiphon           & -0.612   & 1.613    & 0.881   & 0.38            & 0.500      \\
Tor-Obfs4             & 5.237    & 6.630    & 16.688  & \textless{}.001 & 5.934      \\
Snowflake-WebTunnel   & 17.980   & 20.887   & 26.206  & \textless{}.001 & 19.433     \\
Shadowsocks-Meek      & -37.525  & -32.485  & -27.221 & \textless{}.001 & -35.005    \\
Tor-Conjure           & 2.286    & 3.794    & 7.902   & \textless{}.001 & 3.040      \\
Cloak-Conjure         & 2.161    & 3.638    & 7.699   & \textless{}.001 & 2.899      \\
Dnstt-WebTunnel       & 22.298   & 25.979   & 25.699  & \textless{}.001 & 24.138     \\
Tor-Stegotorus        & -14.467  & -8.907   & -8.239  & \textless{}.001 & -11.687    \\
Meek-Psiphon          & 38.246   & 44.357   & 26.493  & \textless{}.001 & 41.301     \\
Stegotorus-Cloak      & 9.155    & 14.581   & 8.574   & \textless{}.001 & 11.868     \\
Tor-Meek              & -42.429  & -37.552  & -32.145 & \textless{}.001 & -39.991    \\
Obfs4-Snowflake       & -22.072  & -19.389  & -30.290 & \textless{}.001 & -20.731    \\
Tor-WebTunnel         & 3.250    & 5.145    & 8.681   & \textless{}.001 & 4.198      \\
Obfs4-Shadowsocks     & -12.766  & -10.818  & -23.728 & \textless{}.001 & -11.792    \\
Obfs4-Meek            & -47.169  & -41.918  & -33.247 & \textless{}.001 & -44.544    \\
Psiphon-Conjure       & -0.169   & 1.971    & 1.651   & 0.1             & 0.901      \\
Cloak-Meek            & -42.907  & -38.108  & -33.090 & \textless{}.001 & -40.507    \\
Marionette-Shadowsocks & 36.119   & 48.980   & 12.969  & \textless{}.001 & 42.549     \\
Marionette-Conjure     & 42.612   & 54.722   & 15.753  & \textless{}.001 & 48.667     \\
Tor-Snowflake         & -16.483  & -14.178  & -26.065 & \textless{}.001 & -15.331    \\
Obfs4-Stegotorus      & -21.663  & -15.173  & -11.123 & \textless{}.001 & -18.418    \\
Snowflake-Conjure     & 17.052   & 19.523   & 29.001  & \textless{}.001 & 18.288     \\
Obfs4-Cloak           & -6.467   & -4.940   & -14.643 & \textless{}.001 & -5.703     \\
Marionette-Stegotorus  & 23.130   & 41.024   & 7.027   & \textless{}.001 & 32.077     \\
Snowflake-Dnstt       & -6.264   & -3.623   & -7.339  & \textless{}.001 & -4.944     \\
Obfs4-Dnstt           & -27.093  & -23.618  & -28.604 & \textless{}.001 & -25.356    \\
Cloak-Psiphon         & -0.658   & 1.444    & 0.733   & 0.46            & 0.393      \\
Marionette-Snowflake   & 19.868   & 35.129   & 7.063   & \textless{}.001 & 27.498     \\
Obfs4-Psiphon         & -6.733   & -4.217   & -8.530  & \textless{}.001 & -5.475     \\ \bottomrule
\end{tabular}
\vspace{2mm}
\caption{Paired t-test results for PT pairs in \Cref{fig:web_access_selenium}---website access using \textbf{selenium} [Part I].}
\label{tab:Fig2b-first}
\end{table*}

\begin{table*}[htbp]
\centering
\begin{tabular}{ccccccc}
\toprule
PT Pair & CI Lower & CI Upper & t-value & \textit{P}-value & Mean Diff. \\
\midrule
Psiphon-WebTunnel      & 1.723   & 4.364   & 4.517   & \textless{}.001 & 3.044   \\
Shadowsocks-Dnstt      & -16.284 & -13.434 & -20.437 & \textless{}.001 & -14.859 \\
Marionette-Dnstt        & 23.225  & 38.892  & 7.771   & \textless{}.001 & 31.059  \\
Tor-Shadowsocks        & -6.860  & -5.168  & -13.936 & \textless{}.001 & -6.014  \\
Meek-Dnstt             & 25.148  & 30.497  & 20.389  & \textless{}.001 & 27.822  \\
Meek-Conjure           & 39.628  & 44.520  & 33.711  & \textless{}.001 & 42.074  \\
Tor-Marionette          & -52.550 & -41.498 & -16.678 & \textless{}.001 & -47.024 \\
Marionette-Meek         & -26.509 & 23.634  & -0.112  & 0.91            & -1.438  \\
Cloak-Snowflake        & -16.281 & -13.880 & -24.617 & \textless{}.001 & -15.081 \\
Dnstt-Conjure          & 21.346  & 24.597  & 27.699  & \textless{}.001 & 22.971  \\
Marionette-Cloak        & 42.501  & 54.202  & 16.198  & \textless{}.001 & 48.352  \\
Shadowsocks-Cloak      & 5.187   & 6.961   & 13.415  & \textless{}.001 & 6.074   \\
Marionette-WebTunnel    & 44.080  & 56.749  & 15.599  & \textless{}.001 & 50.414  \\
Shadowsocks-Psiphon    & 6.409   & 8.867   & 12.184  & \textless{}.001 & 7.638   \\
Cloak-WebTunnel        & 3.209   & 5.130   & 8.509   & \textless{}.001 & 4.170   \\
Stegotorus-Psiphon     & 9.723   & 15.514  & 8.540   & \textless{}.001 & 12.619  \\
Meek-WebTunnel         & 40.695  & 46.267  & 30.589  & \textless{}.001 & 43.481  \\
Stegotorus-Meek        & -65.078 & -36.334 & -6.915  & \textless{}.001 & -50.706 \\
Marionette-Psiphon      & 39.314  & 52.652  & 13.514  & \textless{}.001 & 45.983  \\
Shadowsocks-Stegotorus & -7.521  & -2.640  & -4.080  & \textless{}.001 & -5.080  \\
Snowflake-Psiphon      & 16.650  & 19.977  & 21.576  & \textless{}.001 & 18.313  \\
Conjure-WebTunnel      & 0.541   & 2.276   & 3.181   & 0.002           & 1.408   \\
Obfs4-Marionette        & -57.120 & -46.416 & -18.958 & \textless{}.001 & -51.768 \\
Stegotorus-Conjure     & 10.338  & 16.033  & 9.075   & \textless{}.001 & 13.185  \\
Obfs4-WebTunnel        & -2.385  & -0.565  & -3.175  & 0.002           & -1.475  \\
Snowflake-Meek         & -31.124 & -26.015 & -21.921 & \textless{}.001 & -28.570 \\
Shadowsocks-Conjure    & 8.020   & 9.930   & 18.426  & \textless{}.001 & 8.975  \\
\bottomrule
\end{tabular}
\vspace{2mm}
\caption{Paired t-test results for PT pairs in \Cref{fig:web_access_selenium}---website access using \textbf{selenium} [Part II].}
\label{tab:Fig2b-second}
\end{table*}


\begin{table*}[htbp]
\begin{tabular}{cccccc}
\toprule
PT Pair                 & CI Lower  & CI Upper & t-value & \textit{P}-value         & Mean Diff. \\ \midrule
obfs4-Stegotaurus       & -141.854  & -54.056  & -4.605  & \textless{}.001 & -97.955    \\
obfs4-Shadowsocks       & -201.300  & -26.632  & -2.693  & 0.01            & -113.966   \\
obfs4-Psiphon           & -37.557   & 22.618   & -0.512  & 0.61            & -7.470     \\
obfs4-Tor               & 7.736     & 82.363   & 2.492   & 0.02            & 45.049     \\
obfs4-Cloak             & -21.047   & 77.185   & 1.179   & 0.25            & 28.069     \\
obfs4-Webtunnel         & -17.103   & 77.308   & 1.316   & 0.2             & 30.103     \\
obfs4-Conjure           & -147.839  & 2.570    & -1.993  & 0.06            & -72.635    \\
obfs4-Camoufler         & -136.373  & -39.103  & -3.723  & 0.001           & -87.738    \\
obfs4-Marionette        & -2171.253 & -217.858 & -2.524  & 0.02            & -1194.555  \\
Stegotaurus-Shadowsocks & -96.331   & 64.310   & -0.411  & 0.68            & -16.011    \\
Stegotaurus-Psiphon     & 43.461    & 137.510  & 3.971   & \textless{}.001 & 90.486     \\
Stegotaurus-Tor         & 76.059    & 209.950  & 4.409   & \textless{}.001 & 143.005    \\
Stegotaurus-Cloak       & 52.648    & 199.401  & 3.545   & 0.002           & 126.024    \\
Stegotaurus-Webtunnel   & 55.412    & 200.703  & 3.638   & 0.001           & 128.058    \\
Stegotaurus-Conjure     & -65.413   & 116.054  & 0.576   & 0.57            & 25.321     \\
Stegotaurus-Camoufler   & -38.435   & 58.870   & 0.433   & 0.67            & 10.217     \\
Stegotaurus-Marionette  & -2075.884 & -117.317 & -2.311  & 0.03            & -1096.600  \\
Shadowsocks-Psiphon     & 28.114    & 184.879  & 2.804   & 0.01            & 106.496    \\
Shadowsocks-Tor         & 78.692    & 239.339  & 4.086   & \textless{}.001 & 159.015    \\
Shadowsocks-Cloak       & 58.470    & 225.600  & 3.508   & 0.002           & 142.035    \\
Shadowsocks-Webtunnel   & 67.595    & 220.543  & 3.888   & \textless{}.001 & 144.069    \\
Shadowsocks-Conjure     & -89.598   & 172.261  & 0.652   & 0.52            & 41.331     \\
Shadowsocks-Camoufler   & -32.371   & 84.828   & 0.924   & 0.36            & 26.228     \\
Shadowsocks-Marionette  & -2107.966 & -53.213  & -2.171  & 0.04            & -1080.589  \\
Psiphon-Tor             & 13.818    & 91.220   & 2.801   & 0.01            & 52.519     \\
Psiphon-Cloak           & -18.538   & 89.615   & 1.356   & 0.19            & 35.539     \\
Psiphon-Webtunnel       & -5.903    & 81.048   & 1.784   & 0.09            & 37.572     \\
Psiphon-Conjure         & -154.954  & 24.623   & -1.498  & 0.15            & -65.165    \\
Psiphon-Camoufler       & -124.412  & -36.125  & -3.753  & \textless{}.001 & -80.268    \\
Psiphon-Marionette      & -2174.942 & -199.229 & -2.480  & 0.02            & -1187.086  \\
Tor-Cloak               & -52.687   & 18.726   & -0.982  & 0.34            & -16.980    \\
Tor-Webtunnel           & -39.904   & 10.011   & -1.236  & 0.23            & -14.947    \\
Tor-Conjure             & -209.323  & -26.045  & -2.650  & 0.01            & -117.684   \\
Tor-Camoufler           & -176.706  & -88.869  & -6.240  & \textless{}.001 & -132.787   \\
Tor-Marionette          & -2233.277 & -245.933 & -2.575  & 0.02            & -1239.605  \\
Cloak-Webtunnel         & -27.976   & 32.044   & 0.140   & 0.89            & 2.034      \\
Cloak-Conjure           & -189.178  & -12.229  & -2.349  & 0.03            & -100.704   \\
Cloak-Camoufler         & -158.276  & -73.338  & -5.628  & \textless{}.001 & -115.807   \\
Cloak-Marionette        & -2230.841 & -214.408 & -2.503  & 0.02            & -1222.625  \\
Webtunnel-Conjure       & -196.256  & -9.219   & -2.267  & 0.03            & -102.737   \\
Webtunnel-Camoufler     & -158.284  & -77.397  & -6.014  & \textless{}.001 & -117.840   \\
Webtunnel-Marionette    & -2230.391 & -218.925 & -2.513  & 0.02            & -1224.658  \\
Conjure-Camoufler       & -111.709  & 81.503   & -0.323  & 0.75            & -15.103    \\
Conjure-Marionette      & -2103.733 & -140.109 & -2.358  & 0.03            & -1121.921  \\
Camoufler-Marionette    & -2113.559 & -100.076 & -2.269  & 0.03            & -1106.818  \\ \bottomrule
\end{tabular}
\caption{Paired t-test results for PT pairs in \Cref{fig:file-download-curl}---file download.}
\label{tab:fig5-first}
\end{table*}


\begin{table*}[htbp]
\centering
\begin{tabular}{cccccc}
\toprule
PT Pair               & CI Lower & CI Upper & t-value & \textit{P}-value         & Mean Diff. \\ \midrule
Cloak-Snowflake       & -16.062  & -5.941   & -4.261  & \textless{}.001 & -11.001    \\
Psiphon-WebTunnel     & -0.412   & 0.048    & -1.550  & 0.12            & -0.182     \\
Shadowsocks-Snowflake & -15.137  & -4.830   & -3.797  & \textless{}.001 & -9.984     \\
Cloak-Meek            & -27.588  & -25.223  & -43.763 & \textless{}.001 & -26.405    \\
Obfs4-Marionette       & -60.212  & -31.773  & -6.339  & \textless{}.001 & -45.992    \\
Marionette-WebTunnel   & 34.789   & 64.290   & 6.582   & \textless{}.001 & 49.539     \\
Obfs4-Shadowsocks     & -0.103   & 0.879    & 1.548   & 0.12            & 0.388      \\
Shadowsocks-Psiphon   & 1.752    & 2.497    & 11.170  & \textless{}.001 & 2.125      \\
Meek-Psiphon          & 26.289   & 28.637   & 45.844  & \textless{}.001 & 27.463     \\
Tor-Psiphon           & 0.646    & 1.292    & 5.875   & \textless{}.001 & 0.969      \\
Obfs4-Snowflake       & -16.091  & -7.781   & -5.630  & \textless{}.001 & -11.936    \\
Obfs4-Stegotorus      & -0.840   & 0.119    & -1.472  & 0.14            & -0.360     \\
Obfs4-Psiphon         & 2.026    & 2.898    & 11.063  & \textless{}.001 & 2.462      \\
Tor-Meek              & -27.529  & -25.282  & -46.071 & \textless{}.001 & -26.405    \\
Obfs4-Meek            & -25.847  & -23.645  & -44.043 & \textless{}.001 & -24.746    \\
Snowflake-Psiphon     & 5.083    & 15.395   & 3.892   & \textless{}.001 & 10.239     \\
Tor-Snowflake         & -17.487  & -6.724   & -4.409  & \textless{}.001 & -12.106    \\
Dnstt-WebTunnel       & 6.351    & 8.092    & 16.264  & \textless{}.001 & 7.222      \\
Stegotorus-Meek       & -24.864  & -22.425  & -38.003 & \textless{}.001 & -23.645    \\
Tor-Shadowsocks       & -1.612   & -0.745   & -5.328  & \textless{}.001 & -1.178     \\
Shadowsocks-Conjure   & 2.513    & 3.349    & 13.758  & \textless{}.001 & 2.931      \\
Tor-Stegotorus        & -2.483   & -1.604   & -9.105  & \textless{}.001 & -2.043     \\
Tor-Obfs4             & -2.006   & -1.254   & -8.489  & \textless{}.001 & -1.630     \\
Marionette-Snowflake   & 17.386   & 49.685   & 4.070   & 0.001           & 33.536     \\
Stegotorus-Snowflake  & -17.443  & -4.310   & -3.246  & 0.004           & -10.877    \\
Marionette-Conjure     & 35.787   & 65.260   & 6.720   & \textless{}.001 & 50.524     \\
Marionette-Cloak       & 34.489   & 64.707   & 6.434   & \textless{}.001 & 49.598     \\
Snowflake-WebTunnel   & 8.806    & 16.293   & 6.570   & \textless{}.001 & 12.549     \\
Cloak-Conjure         & 1.133    & 1.897    & 7.764   & \textless{}.001 & 1.515      \\
Shadowsocks-Cloak     & 1.195    & 1.967    & 8.022   & \textless{}.001 & 1.581      \\
Stegotorus-Cloak      & 1.776    & 2.867    & 8.347   & \textless{}.001 & 2.321      \\
Stegotorus-Conjure    & 3.312    & 4.177    & 16.972  & \textless{}.001 & 3.745      \\
Meek-Conjure          & 27.038   & 29.396   & 46.901  & \textless{}.001 & 28.217     \\
Snowflake-Dnstt       & 5.787    & 17.650   & 3.872   & \textless{}.001 & 11.719     \\
Marionette-Dnstt       & 27.386   & 60.823   & 5.170   & \textless{}.001 & 44.105     \\
Meek-WebTunnel        & 25.903   & 28.243   & 45.353  & \textless{}.001 & 27.073     \\
Dnstt-Conjure         & 7.463    & 9.173    & 19.073  & \textless{}.001 & 8.318      \\
Tor-WebTunnel         & 0.416    & 0.989    & 4.812   & \textless{}.001 & 0.702      \\
Obfs4-Conjure         & 2.894    & 3.694    & 16.137  & \textless{}.001 & 3.294      \\ \bottomrule
\end{tabular}
\vspace{2mm}
\caption{Paired t-test results for PT pairs in \Cref{fig:speed_index}---speed index [Part I].}
\label{tab:Fig11-first}
\end{table*}

\begin{table*}[htbp]
\centering
\begin{tabular}{ccccccc}
\toprule
PT Pair & CI Lower & CI Upper & t-value & \textit{P}-value & Mean Diff. \\
\midrule
Dnstt-Psiphon          & 6.447   & 8.193   & 16.433  & \textless{}.001 & 7.320   \\
Shadowsocks-Stegotorus & -1.075  & -0.198  & -2.845  & 0.005           & -0.636  \\
Tor-Dnstt              & -7.188  & -5.545  & -15.193 & \textless{}.001 & -6.366  \\
Obfs4-Cloak            & 1.372   & 2.262   & 8.001   & \textless{}.001 & 1.817   \\
Marionette-Psiphon      & 33.461  & 63.266  & 6.361   & \textless{}.001 & 48.363  \\
Obfs4-Dnstt            & -5.542  & -3.986  & -12.004 & \textless{}.001 & -4.764  \\
Tor-Cloak              & -0.319  & 0.408   & 0.239   & 0.81            & 0.044   \\
Stegotorus-Dnstt       & -4.908  & -3.100  & -8.682  & \textless{}.001 & -4.004  \\
Cloak-Psiphon          & 0.671   & 1.211   & 6.824   & \textless{}.001 & 0.941   \\
Stegotorus-WebTunnel   & 2.257   & 3.240   & 10.968  & \textless{}.001 & 2.749   \\
Shadowsocks-Meek       & -26.274 & -24.006 & -43.451 & \textless{}.001 & -25.140 \\
Tor-Conjure            & 1.394   & 1.925   & 12.235  & \textless{}.001 & 1.659   \\
Cloak-Dnstt            & -8.345  & -6.532  & -16.083 & \textless{}.001 & -7.439  \\
Snowflake-Meek         & -18.355 & -8.839  & -5.601  & \textless{}.001 & -13.597 \\
Shadowsocks-Dnstt      & -5.921  & -4.236  & -11.816 & \textless{}.001 & -5.078  \\
Psiphon-Conjure        & 0.449   & 1.136   & 4.517   & \textless{}.001 & 0.793   \\
Marionette-Shadowsocks  & 34.535  & 62.909  & 6.731   & \textless{}.001 & 48.722  \\
Marionette-Stegotorus   & 31.222  & 62.641  & 5.855   & \textless{}.001 & 46.931  \\
Cloak-WebTunnel        & 0.324   & 0.892   & 4.192   & \textless{}.001 & 0.608   \\
Conjure-WebTunnel      & -1.216  & -0.568  & -5.397  & \textless{}.001 & -0.892  \\
Meek-Dnstt             & 19.315  & 21.379  & 38.653  & \textless{}.001 & 20.347  \\
Tor-Marionette          & -60.658 & -30.748 & -5.990  & \textless{}.001 & -45.703 \\
Obfs4-WebTunnel        & 1.965   & 2.774   & 11.477  & \textless{}.001 & 2.370   \\
Shadowsocks-WebTunnel  & 1.647   & 2.395   & 10.589  & \textless{}.001 & 2.021   \\
Marionette-Meek         & 2.987   & 35.628  & 2.319   & 0.03            & 19.307  \\
Snowflake-Conjure      & 10.653  & 22.271  & 5.555   & \textless{}.001 & 16.462  \\
Stegotorus-Psiphon     & 2.285   & 3.362   & 10.281  & \textless{}.001 & 2.824 \\
\bottomrule
\end{tabular}
\vspace{2mm}
\caption{Paired t-test results for PT pairs in \Cref{fig:speed_index}---speed index [Part II].}
\label{tab:Fig11-second}
\end{table*}

\begin{table*}[htbp]
\centering
\begin{tabular}{cccccc}
\toprule
PT Category Pair                     & CI Upper & CI Lower & t-value & \textit{P}-value         & Mean Diff. \\ \midrule
fully encrypted-mimicry     & -5.481   & -4.946   & -38.256 & \textless{}.001 & -5.214     \\
mimicry-Tor                 & 3.974    & 4.557    & 28.692  & \textless{}.001 & 4.265      \\
proxy layer-Tor             & 0.809    & 1.229    & 9.507   & \textless{}.001 & 1.019      \\
Tor-tunneling               & -4.211   & -3.581   & -24.232 & \textless{}.001 & -3.896     \\
mimicry-tunneling           & 0.019    & 0.692    & 2.069   & 0.04            & 0.355      \\
fully encrypted-proxy-layer & -2.191   & -1.730   & -16.643 & \textless{}.001 & -1.960     \\
fully encrypted-tunneling   & -5.234   & -4.597   & -30.266 & \textless{}.001 & -4.915     \\
mimicry-proxy layer         & 2.974    & 3.490    & 24.554  & \textless{}.001 & 3.232      \\
fully encrypted-Tor         & -1.239   & -0.650   & -6.279  & \textless{}.001 & -0.944     \\
proxy layer-tunneling       & -3.167   & -2.607   & -20.235 & \textless{}.001 & -2.887     \\ \bottomrule
\end{tabular}
\caption{Paired t-test results for PT category pairs in \Cref{fig:web_access}---website access using \textbf{curl}.}
\label{tab:Fig2a-PT-category}
\end{table*}

\end{document}